\documentclass[twocolumn]{aastex6}
\pdfoutput=1
\usepackage{graphicx}
\usepackage[version=4]{mhchem}
\usepackage{amssymb,amsfonts,amsmath,amstext,amsgen,amsopn,amsxtra,indentfirst,times,makecell}
\usepackage{soul} 
\newcounter{reactions} 
\newenvironment{reactions}{\refstepcounter{reactions}\equation}{\tag{R\thereactions}\endequation}
\usepackage{comment}
\begin{document}

\title{Towards Consistent Modeling of Atmospheric Chemistry and Dynamics in Exoplanets: Validation and Generalization of Chemical Relaxation Method}

\author{Shang-Min Tsai\altaffilmark{1}, Daniel Kitzmann\altaffilmark{1}, James R. Lyons\altaffilmark{2}, Jo\~ao Mendon\c ca\altaffilmark{1,3,4}, Simon L. Grimm\altaffilmark{1}, Kevin Heng\altaffilmark{1}}
\altaffiltext{1}{University of Bern, Center for Space and Habitability, Gesellschaftsstrasse 6, CH-3012, Bern, Switzerland.  Email: shang-min.tsai@space.unibe.ch, kevin.heng@csh.unibe.ch}
\altaffiltext{2}{Arizona State University, School of Earth \& Space Exploration, Tempe, Arizona 85281 U.S.A.}
\altaffiltext{3}{University of Copenhagen, Centre for Star and Planet Formation, Niels Bohr Institute and Natural History Museum of Denmark, Ostervoldgade 5--7, 1350 Copenhagen K, Denmark}
\altaffiltext{4}{Astrophysics and Atmospheric Physics, National Space Institute, Technical University of Denmark, Elektrovej, 2800, Kgs. Lyngby, Denmark}

\begin{abstract}
Motivated by the work of Cooper \& Showman, we revisit the chemical relaxation method, which seeks to enhance the computational efficiency of chemical-kinetics calculations by replacing the chemical network with a handful of independent source/sink terms. Chemical relaxation solves the evolution of the system and can treat disequilibrium chemistry, as the source/sink terms are driven towards chemical equilibrium on a prescribed chemical timescale, but it has surprisingly never been validated.  First, we generalize the treatment by forgoing the use of a single chemical timescale, instead developing a pathway analysis tool that allows us to identify the rate-limiting reaction as a function of temperature and pressure.  For the interconversion between methane and carbon monoxide and between ammonia, and molecular nitrogen, we identify the key rate-limiting reactions for conditions relevant to currently characterizable exo-atmospheres (500--3000 K, 0.1 mbar to 1 kbar).  Second, we extend chemical relaxation to include carbon dioxide and water.  Third, we examine the role of metallicity and carbon-to-oxygen ratio in chemical relaxation.  Fourth, we apply our pathway analysis tool to diagnose the differences between our chemical network and that of Moses and Venot.  Finally, we validate the chemical relaxation method against full chemical kinetics calculations in one dimension.  For WASP-18b-, HD 189733b- and GJ 1214-b-like atmospheres, we show that chemical relaxation is mostly accurate to within an order of magnitude, a factor of 2 and $\sim 10\%$, respectively. The level of accuracy attained allows for the chemical relaxation method to be included in three-dimensional general circulation models.
\end{abstract}

\keywords{planets and satellites: atmospheres -- planets and satellites: composition -- methods: numerical}

\section{Introduction}

The study of exoplanets has evolved from detection to characterization, thanks to the advent of cutting-edge observational techniques.  The spectra of exo-atmospheres provide us with valuable clues about the atmospheric chemistry and thermal structure.  Diagnosing and interpreting these spectra to obtain chemical compositions is now at the forefront of exo-atmospheric research.  

The simplest assumption is to build a model in chemical equilibrium, where the molecular composition for a given elemental abundance only depends on local, basic parameters (pressure and temperature) independent of the reaction pathways. However, equilibrium chemistry only holds in the hot ($T \gtrsim 2000$ K) or deep ($P \gtrsim 100$ bar) parts of the atmosphere.  Processes like ultraviolet (UV) irradiation and atmospheric dynamics drive the chemical composition in the observable atmosphere away from equilibrium. These disequilibrium processes commonly dominate the observable parts of atmospheres in the Solar System.

Chemical kinetics models, which incorporate a chemical network of hundreds to thousands of reactions, are needed to study disequilibrium chemistry (e.g., \citealt{koppa12,moses11,venot12,krome14,venot12,hu14,rm16,tsai17}).  Even with such a number of reactions, chemical networks are greatly reduced compared to Nature and restricted to measurements in specific temperature ranges.  In other words, there does not exist a one-fits-all chemical network in practice.  The accuracy of a network is determined by whether it includes the relevant reactions and if the input rate coefficients are reliable. Unfortunately, these chemical kinetics models are expensive to run, especially when one desires to couple chemistry to three-dimensional atmospheric motion.  In one dimension, eddy diffusion is used to mimic large-scale atmospheric circulation, convection, turbulence, etc.  The three-dimensional atmospheric circulation patterns of tidally-locked, highly-irradiated exoplanets are demonstrably more complicated (e.g., \citealt{showman09,dobbs13} and see \citealt{hs15} for a review).  In order to correctly interpret the transmission spectra of hot Jupiters, there is a need to develop three-dimensional general circulation models (GCMs) that couple the atmospheric dynamics and chemistry.  Intermediate steps have already been taken in this direction by, e.g., \cite{agu12,agu14}, who coupled a chemical kinetics code to a simplified model of the atmospheric dynamics (constant solid-body rotation mimicking a uniform equatorial jet). There is clearly a need to build on studies like these, but a brute-force coupling between a three-dimensional solver of the fluid equations and a chemical kinetics code with a network of hundreds to thousands of reactions is computationally challenging, even without considering radiative transfer (which is needed to include photochemistry).

Another approach is to simplify the chemical scheme. Conceptually, the interaction between atmospheric motion and chemistry is a comparison between two timescales: the dynamical versus chemical timescales.  The simplest approach is to use the timescale argument, or quenching approximation.  It assumes that the deep atmosphere is in chemical equilibrium, because the chemical timescale is much shorter than the dynamical timescale.  There is a location within the atmosphere known as the quench point, where the timescales are equal.  Above the quench point, the chemical abundances are assumed to be well-mixed and frozen to their equilibrium values at the quench point. In technical parlance, the process is referred as ``transport-induced quenching". The quenching approximation has been used to understand the over-abundance of carbon monoxide (CO) in the upper troposphere of Jupiter \citep{prinn}.  \cite{vc12} compared the chemical timescale for converting methane (CH$_4$) to CO to the orbital timescales of highly eccentric exoplanets to study the interaction between the evolving thermal structures and the chemistry.  While the quenching approximation can be used to build the bulk of our intuition, it has been shown that it should be applied with caution to know when the abundance of a molecule is controlled by the disequilibrium abundance of another parent molecule, e.g., acetylene being controlled by methane \citep{tsai17}. Additionally, the quenching approximation contains the ambiguity of having to specify an appropriate length scale, which is not known from first principles \citep{smith98}.

A more realistic approach is chemical relaxation. It was pioneered in the exo-atmospheres literature by \cite{cs06}.  Instead of simplifying the treatment of atmospheric dynamics, chemical relaxation takes the approach of replacing the chemical network with a single source/sink term that depends on the chemical timescale.  \cite{cs06} coupled chemical relaxation to a simplified GCM to study quenching of CO, and suggested that most of the carbon is locked up in CO in HD 209458b due to transport. However, a shortcoming of their study is the assumption of a single rate-limiting reaction for the interconversion and probably the underestimation of the timescale of CO.

\begin{figure}
\begin{center}
\vspace{-0.1in}
\includegraphics[width=\columnwidth]{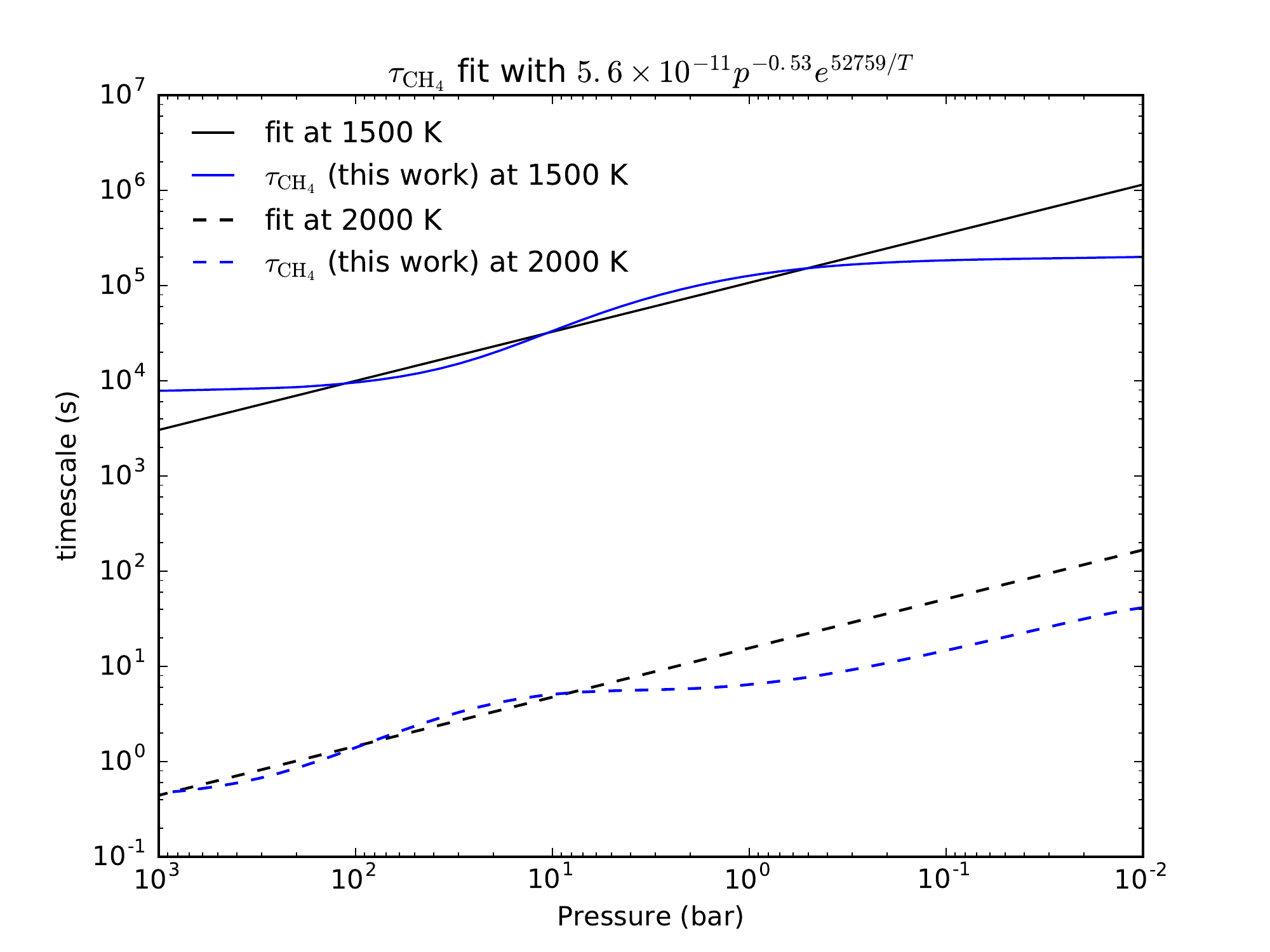}
\end{center}
\vspace{-0.2in}
\caption{Chemical timescales associated with methane for 1500 K (blue solid curve) and 2000 K (blue dashed curve) computed using full chemical kinetics.  Following Figure 5 of \cite{zm14} $^1$, we found a fit for $\tau_{\rm CH_4}$ in the range of $1300 \mbox{ K} \lesssim T \lesssim 2200$ K and $0 \lesssim P \lesssim 100$ bar: $5.6 \times 10^{-11} p^{-0.53} e^{52759 \mbox{ (K)}/T}$.  We then overlaid this Arrhenius-like fit (black, solid and dashed curves) with full chemical-kinetics calculations of $\tau_{\rm CH_4}$, over a broader range of pressures, in  for two values of the temperature (1500 and 2000 K).}
\label{fig:fit_ZM}
\end{figure}

A prerequisite for implementing chemical relaxation is to be able to compute the chemical timescale over a broad range of temperatures and pressures.  Initially, we had hoped to seek a universal fit for the chemical timescale across temperature, pressure and metallicity, motivated by the work of \cite{zm14}.  Figure \ref{fig:fit_ZM} shows our attempt at fitting an Arrhenius-like function to the chemical timescale for CH$_4$-CO conversion. We find the timescales explored in \cite{zm14} are likely restricted to a narrow temperature-pressure range. It is apparent that such a simple approach fails to fit the timescale across the range of values of temperature and pressure needed for us to implement chemical relaxation.

\footnotetext[1]{Note that $\tau_{\ce{CO}}$ is not distinguished from $\tau_{\ce{CH4}}$ in \cite{zm14}: these two timescales are only equal when [CO]/[\ce{CH4}] = 1. We find in fact the left panel in Figure 5 of \cite{zm14} shows $\tau_{\ce{CH4}}$ and the right panel shows $\tau_{\ce{CO}}$ due to different temperatures.}

Generally, the chemical timescale is a \textit{function} that draws upon different rate-limiting reactions in different temperature and pressure regimes. A challenge with the chemical relaxation method is to find these rate-limiting reactions. In doing so, we develop a simple method to identify the dominant pathway and the associated rate-limiting step (RLS) in the chemical network for given values of the temperature, pressure, and elemental abundances. The pathway analysis can be used for comparing chemical kinetics calculations performed by different groups using different chemical networks, as it allows one to identify the key reactions that essentially control the output, and assess if key reactions are missing. It is useful both on from a practical point of view and for developing physical intuition (e.g., \citealt{thomas}).
 
Ever since the work of \cite{cs06}, only \cite{ben18} has recently implemented the relaxation method in the Met Office Unified Model and improved it by consistently coupling to radiative transfer. However, to our knowledge, chemical relaxation has surprisingly never been validated. By ``validation", we mean that the accuracy of chemical relaxation should be demonstrated to a factor of a few, rather than to $\sim 1\%$ accuracy (or better), given the existing uncertainties in the rate coefficients and the approximate nature of the approach. In addition, transit radii are proportional to the logarithms of the chemical abundances, such a factor-of-several validation suffices for studying atmospheric chemistry in hot Jupiters.  In the current study, we perform this validation step in one dimension, since it is a test of the ability of the chemical relaxation scheme to mimic the full chemical network, rather than one of the complicated three-dimensional geometry.  In a future work, we will aim to couple the chemical relaxation scheme to a three-dimensional GCM, but in the current work we restrict ourselves to validating the scheme on a factor-of-several basis.  Such a strategy follows the well-established hierarchical approach of constructing climate models \citep{held05}.  We do not consider photochemistry for the current study.

In \S\ref{sec:theory}, we provide the background theory on how to compute the chemical timescale.  In \S\ref{sec:rls}, we describe our methodology for identifying the rate-limiting chemical reactions.  In \S\ref{sec:chemtime}, we describe how we compute the chemical timescales.  In \S\ref{sec:pathway}, we compare our chemical network to that of Moses and Venot using our pathway analysis tool.  In \S\ref{sec:validate}, we validate the chemical relaxation method for three model atmospheres.  In \S\ref{sec:summary}, we summarize our findings and list opportunities for future work.

\section{Simplified expression for reaction rate equations}
\label{sec:theory}

To compute the chemical timescale for use in the chemical relaxation method, we need to first derive a simplified expression for it that depends only on the local conditions of temperature and  pressure.  This allows us to evaluate the rate of change of the abundances efficiently.

Consider a chemical species subject to production (${\cal P}$)  and loss (${\cal L}$) through a network of chemical reactions without any disequilibrium process (e.g., transport).  The rate of change of its volume number density ($n$) is
\begin{equation}
\frac{d n}{d t} = {\cal P} - n {\cal L}.
\label{eq:master}
\end{equation}
Equation (\ref{eq:master}) is written in a way that ${\cal P}$ and ${\cal L}$ do not depend on $n$, as production only depends on other species and loss depends linearly on the number density of the reactant for a typical bimolecular reaction. Since ${\cal P}$ and ${\cal L}$ depend on the number densities of other species, the ensemble of equation (\ref{eq:master}) for every species forms a system of coupled differential equations and the calculation involves inverting a large matrix (e.g., \citealt{hu12,tsai17}). If one wishes to implement chemical kinetics into the dynamical core of a GCM, then one needs to include a separate Euler equation for every species in the chemical network.

The relaxation method rewrites equation (\ref{eq:master}) as \citep{smith98,cs06}
\begin{equation}
\frac{dn}{dt} = -\frac{n-n_\mathrm{EQ}}{\tau_\mathrm{chem}},
\label{eq:relax}
\end{equation}
replacing the production and loss terms with a source/sink term that relaxes $n$ to $n_\mathrm{EQ}$ by the chemical timescale ($\tau_{\rm chem}$). When the abundance is greater (less) than its equilibrium value, then it decreases (increases) and works its way towards equilibrium. The chemical timescale is effectively determined by the employed chemical network. When coupled to a GCM, whether the species in question attains chemical equilibrium depends on the competition between atmospheric dynamics and chemistry via their corresponding timescales.  Chemical relaxation is analogous to the Newtonian relaxation method used as a substitute for radiative transfer (e.g., \citealt{hs94}) or the treatment of condensation where the supersaturated gas is relaxed to the saturated vapor number density on the condensation timescale \citep{hu12}.

The chemical timescale is conventionally expressed without justification as
\begin{equation}
\tau_\mathrm{chem} = \frac{n}{\vert dn/dt \vert} = \frac{n_\mathrm{EQ}}{n^\prime_\mathrm{EQ} {\cal L}^\prime_\mathrm{EQ}},
\label{eq:tau}
\end{equation}
where the second equality assumes that the species in question has a number density that is equal to its chemical equilibrium value.  Furthermore, $n^\prime_\mathrm{EQ} {\cal L}^\prime_\mathrm{EQ}$ refers to the loss rate determined by the RLS involving other species but not $n$, denoted with primes (The ambiguity comes from using the equilibrium abundance for the numerator but not the denominator in equation(\ref{eq:tau}) since $dn/dt$ vanishes in chemical equilibrium).

We now wish to demonstrate that equation (\ref{eq:tau}) may be derived from equation (\ref{eq:master}) and (\ref{eq:relax}).  First, consider the situation when $n \ll n_{\rm EQ}$.  This implies that the loss of the species being considered is negligible compared to production, which means $dn/dt \approx {\cal P}$.  If we assume that all of the other species in the network are close to chemical equilibrium (as most of the intermediate species are fast-reacting radicals), then we can further write ${\cal P} \approx {\cal P}^\prime_{\rm EQ}$ because recall that ${\cal P}$ does not depend on $n$.  For the RLS, we can write ${\cal P}^\prime_{\rm EQ} = n^\prime_{\rm EQ} {\cal L}^\prime_{\rm EQ}$.  By inserting this expression into equation (\ref{eq:master}) and (\ref{eq:relax}), we obtain equation (\ref{eq:tau}).  In the opposite limit of $n \gg n_{\rm EQ}$, the loss of the species dominates production and we have $dn/dt = -n {\cal L} \approx - n^\prime_{\rm EQ} {\cal L}^\prime_{\rm EQ}$.  This again leads to equation (\ref{eq:tau}) from equation (\ref{eq:master}) and (\ref{eq:relax}). Since the chemical timescale expression is approximately correct for both limits, it is expected to work at order-of-magnitude accuracy  at least. This expectation will be confirmed by full numerical calculations of chemical kinetics.

\section{Determining the rate-limiting reactions in the chemical network}
\label{sec:rls}
\begin{figure}
\begin{center}
\vspace{-0.1in}
\includegraphics[width=\columnwidth]{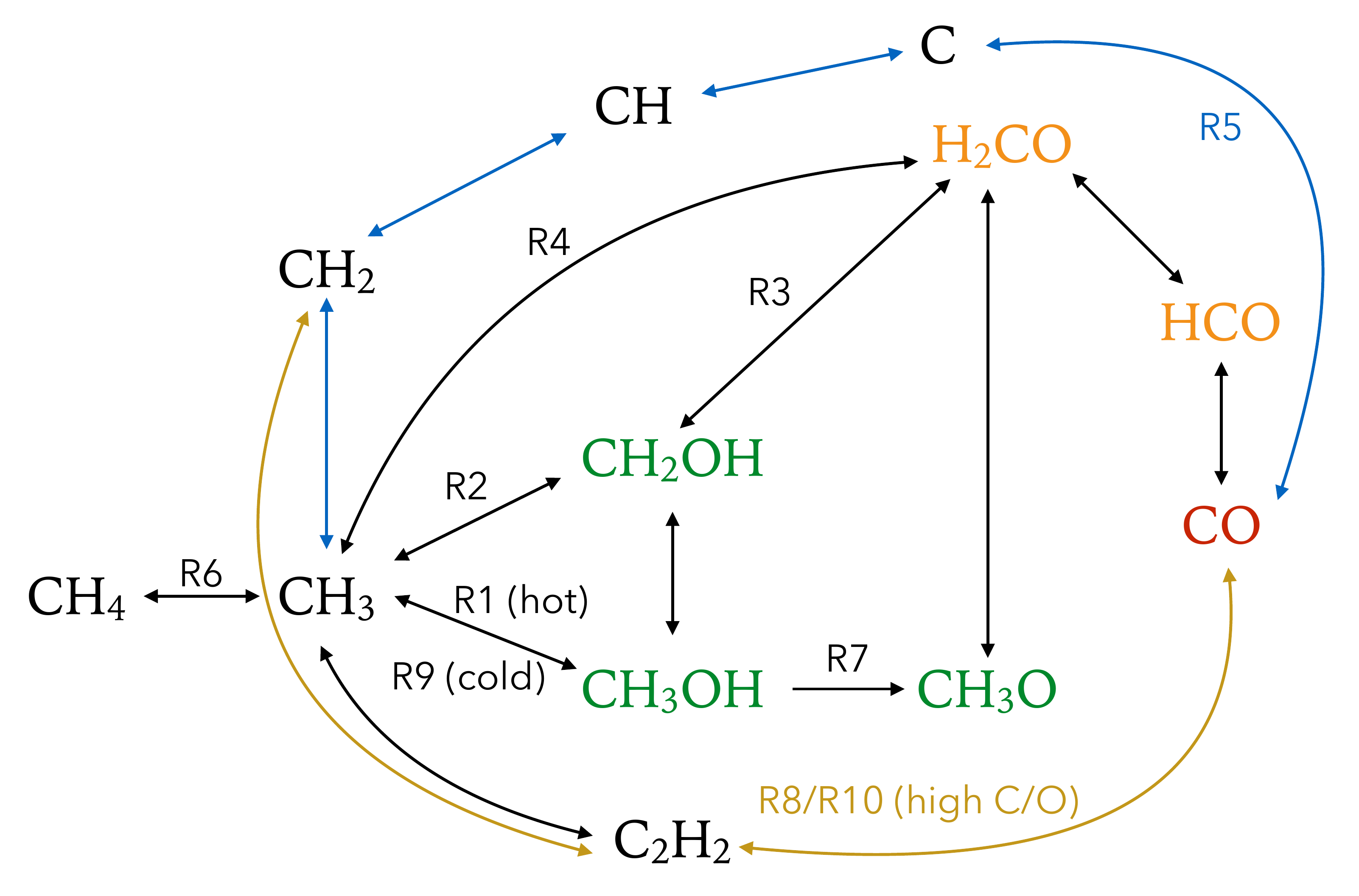}
\end{center}
\vspace{-0.2in}
\caption{Visualization of the major chemical pathways between methane and carbon monoxide in hydrogen-dominated atmospheres.  The triple, double and single bonds between carbon and oxygen are colored red, orange and green, respectively.  The blue arrows represent the pathway at high temperatures and low pressures for $\mbox{C/O} < 1$.  The brown arrows represent the pathway turned up for $\mbox{C/O} > 1$.  For a description of the (\ref{R1}) to (\ref{R10}) reactions, see text. For their specific operating temperatures and pressures, see Tables \ref{tab: pathways} (solar abundance) and \ref{tab: pathways_CtoO2} (C/O=2).}
\label{fig:scheme}
\end{figure}

It is common for the chemical conversion of one species to another to not occur in one step. Rather, it takes multiple steps to surmount the energy barrier via the breaking or forming of chemical bonds.  These sequence of reactions form a pathway, and the chemical timescales associated with each step in the pathway may differ by many orders of magnitude.  The efficiency of a pathway is bottlenecked by its slowest reaction.  Specifically, \textit{the RLS is defined as the slowest reaction along the fastest pathway.}  It informs the effective loss rate, $n^\prime_\mathrm{EQ} {\cal L}^\prime_\mathrm{EQ}$, in equation (\ref{eq:tau}).  Thus, computing the chemical timescale involves identifying the RLS.  For example, \cite{zm14} have remarked how the conversion of CO to CH$_4$ may be visualized as the reduction of the bond between C and O from a triple bond to a double bond to a single bond and eventually splitting C from O, in three steps: first between CO and formaldehyde (H$_2$CO), second between formaldehyde and methanol (CH$_3$OH) and finally between methanol and methane. We build upon and extend the diagram in Figure \ref{fig:scheme}, where the temperature-and-pressure dependent pathways and RLSs are included, as explained in the following subsections.

\begin{figure}
\begin{center}
\vspace{-0.1in}
\includegraphics[width=\columnwidth]{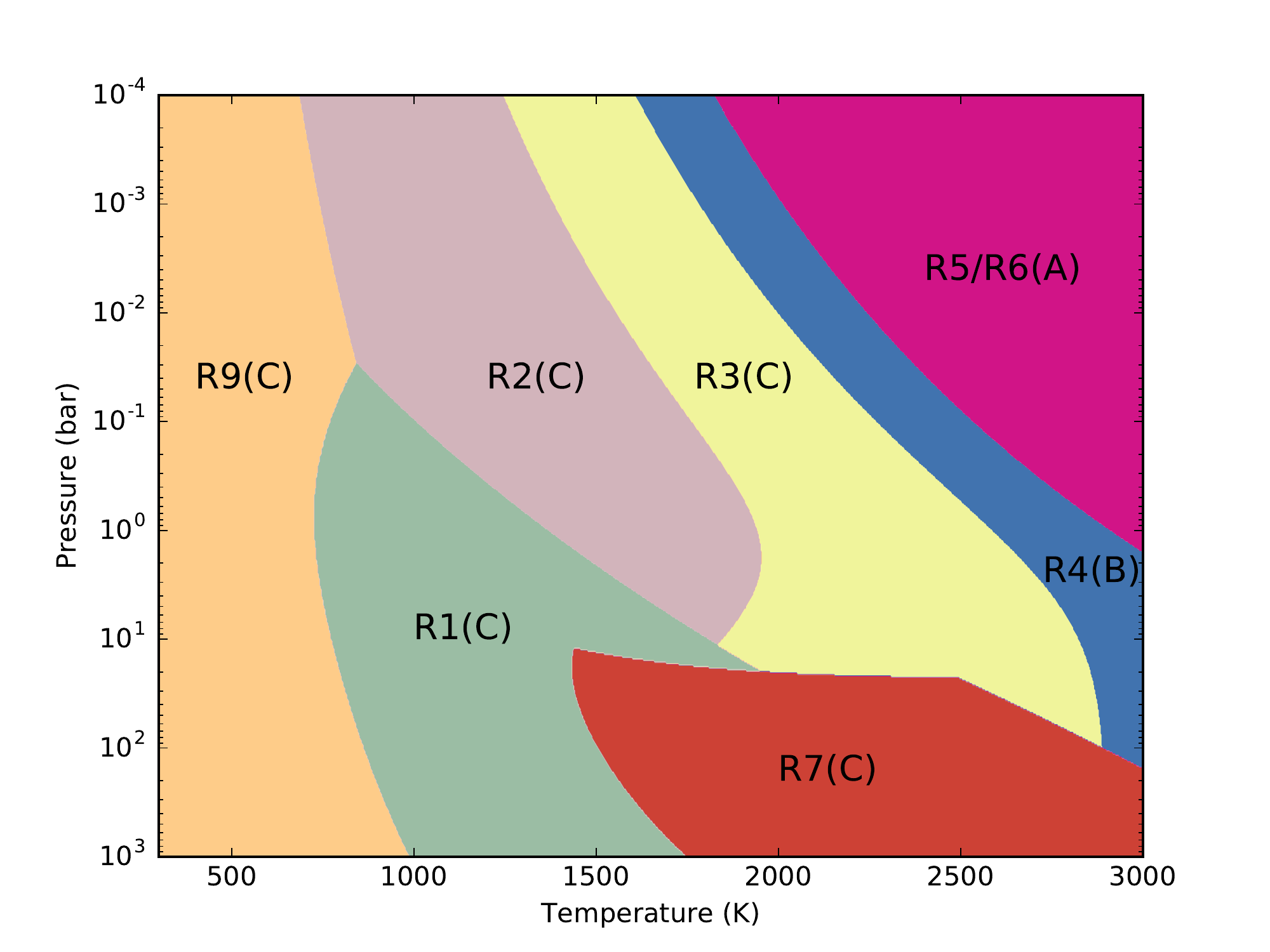}
\end{center}
\vspace{-0.2in}
\caption{Parameter space of temperature and pressure showing how the rate-limiting step corresponds to different chemical reactions (R1 to R9; for more details, see text).  These 9 chemical reactions may in turn be visualized as belonging to 3 different schemes (A, B and C; see text).}
\label{fig: rls}
\end{figure}

\subsection{CH$_4$-CO interconversion}

\subsubsection{Identifying the rate-limiting steps from full chemical kinetics}

In addition to the RLS being a function of temperature and pressure, extracting this information to identify it is not straightforward because the possible number of paths grow exponentially with increasing number of species, making it difficult to track them all. Our approach is to develop a tool using Dijkstra's algorithm to find the shortest path and the associated RLS; see Appendix \ref{app:pathway} for more details.  

Figure \ref{fig: rls} shows a survey of the different RLSs as functions of temperature and pressure and for protosolar elemental abundances from \citet{lodders09} ($\mbox{C/H}=2.776 \times 10^{-4}, \mbox{O/H}=6.062 \times 10^{-4}, \mbox{N/H}=8.185 \times 10^{-5}, \mbox{He/H} = 9.69 \times 10^{-2}$).  Unsurprisingly, there does not exist a single RLS for the range of temperatures (500--3000 K) and pressures (0.1 mbar--1 kbar) considered. The different reactions, labeled R1 to R10, are
\begin{reactions}\label{R1}
\ce{OH + CH3 + M -> CH3OH + M} 
\end{reactions}
\begin{reactions}\label{R2}
\ce{OH + CH3 -> CH2OH + H}
\end{reactions}
\begin{reactions}\label{R3}
\ce{CH2OH + M -> H + H2CO + M}
\end{reactions}
\begin{reactions}\label{R4}
\ce{CH3 + O -> H2CO + H}
\end{reactions}
\begin{reactions}\label{R5}
\ce{OH + C -> CO + H}
\end{reactions}
\begin{reactions}\label{R6}
\ce{H + CH4 -> CH3 + H2}
\end{reactions}
\begin{reactions}\label{R7}
\ce{CH3OH + H -> CH3O + H2}
\end{reactions}
\begin{reactions}\label{R8}
\ce{C2H2 + O -> CH2 + CO}
\end{reactions}
\begin{reactions}\label{R9}
\ce{CH3 + H2O -> CH3OH + H}
\end{reactions}
\begin{reactions}\label{R10}
\ce{C2H2 + OH -> CH3 + CO}
\end{reactions}
where M refers to any third body (i.e. the total number density of the gas). Reaction R8 and R10 are only relevant when $\mbox{C/O}>1$, as we will describe shortly in the example of $\mbox{C/O}=2$.

\subsubsection{Grouping of reactions into three schemes}

It is possible to understand CH$_4$-CO interconversion as consisting of three schemes (at least, for solar-like elemental abundances). As temperature increases, the scheme moves from (C) to (A), as higher kinetic energy allows more ambitious steps.:
\begin{itemize}

\item (A): Though the progressive dehydrogenation of CH$_4$ into C, followed by oxidization into CO (blue arrows in Fig \ref{fig:scheme});

\item (B): Via H$_2$CO (formaldehyde) from directly oxidizing CH$_3$ as described by (\ref{R4});

\item (C): Via intermediate species like CH$_2$OH, CH$_3$OH, or CH$_3$O from oxidizing CH$_3$ (through the molecules shown in green in Fig \ref{fig:scheme}).

\end{itemize}
At high temperatures and low pressures (the magenta region in Figure \ref{fig: rls}), scheme (A) is favored because it requires a high abundance of atomic hydrogen, produced mainly by thermal decomposition of H$_2$. Scheme (B) sits in the transition between scheme (A) and (C).  Scheme (C) covers the broadest range of temperature and pressure and contains the pathways previously identified by, e.g., \cite{yung88,bezard,moses11,vc12}. As shown in Figure \ref{fig:scheme}, CH$_4$ is first converted to CH$_3$ before being oxidized by OH or H$_2$O to form CH$_2$OH or CH$_3$OH/CH$_3$O depending on the temperature and pressure.  These singly-bonded (\ce{C-O}) intermediate species make forming the double bond (\ce{C=O}) in H$_2$CO easier than directly from C and O.  H$_2$CO goes on to efficiently produce HCO and finally the triple-bonded structure of CO.  At high enough temperatures ($T \gtrsim 2000$ K), there is sufficient energy to directly form the double bond between C and O into H$_2$CO via reaction (\ref{R4}) without passing through the intermediate species, which is scheme (B).  This is similar to the RLS, \ce{CH3 + OH -> H2 + H2CO}, initially suggested by \cite{prinn} to explain the quenched CO found in Jupiter.  At even higher temperatures, if the pressure remains low enough then scheme (A) dominates.  Molecular hydrogen is dissociated into atomic hydrogen, which in turn promotes the dehydrogenation of methane.  Eventually, the accumulated C is present at high enough abundances that allow for its oxidization into CO.

A typical pathway of scheme (A) is:
\begin{subequations}
\label{re:A}
\begin{align}
\begin{split} 
\ce{ CH4 + H} &\rightarrow \ce{ CH3 + H2 } \\
\ce{ CH3 + H} &\rightarrow \ce{ CH2 + H2  } \\
\ce{ CH2 + H} &\rightarrow \ce{ CH + H2  } \\
\ce{ CH + H} &\rightarrow \ce{ C + H2 } \\
\ce{ C + OH} &\rightarrow \ce{ CO + H  } \\
\ce{ H + H2O}&\rightarrow \ce{ OH + H2 } \\
\ce{ 2( H2 + M}&\rightarrow \ce{ 2H + M )} \\
\hline \nonumber
\mbox{net} : \ce{ CH4 + H2O} &\rightarrow \ce{CO + 3H2 } 
\end{split} 
\end{align}
\end{subequations}

Scheme (B), which does not involve intermediate species like \ce{CH3OH} or \ce{CH2OH}, goes through the pathway:
\begin{subequations}
\label{re:B}
\begin{align}
\begin{split} 
\ce{ CH4 + H} &\rightarrow \ce{ CH3 + H2 } \\
\ce{ CH3 + O} &\rightarrow \ce{ H2CO + H  } \\
\ce{ H2CO + H} &\rightarrow \ce{ HCO + H2 } \\
\ce{ HCO + H} &\rightarrow \ce{ CO + H2 } \\
\ce{ OH + H} &\rightarrow \ce{ O + H2 } \\
\ce{ H + H2O }&\rightarrow \ce{ OH + H2 } \\
\ce{ 2( H2 + M}&\rightarrow \ce{ 2H + M )} \\
\hline \nonumber
\mbox{net} : \ce{ CH4 + H2O} &\rightarrow \ce{CO + 3H2 } 
\end{split} 
\end{align}
\end{subequations}

Scheme (C) includes several main pathways for $T$ $\lesssim$ 2000 K or P $\gtrsim$ 1 bar. One example of scheme (C) is:
\begin{subequations}
\label{re:C}
\begin{align}
\begin{split} 
\ce{ CH4 + H} &\rightarrow \ce{ CH3 + H2 } \\
\ce{ CH3 + OH + M}&\rightarrow \ce{ CH3OH + M  } \\
\ce{ CH3OH + H }&\rightarrow \ce{ CH3O + H2 } \\
\ce{ CH3O + M }&\rightarrow \ce{ H2CO + H + M } \\
\ce{ H2CO + H }&\rightarrow \ce{ HCO + H2 } \\
\ce{ HCO + M }&\rightarrow \ce{ H + CO + M } \\
\ce{ H + H2O }&\rightarrow \ce{ OH + H2 } \\
\ce{ H2 + M }&\rightarrow \ce{ 2H + M }\\
\hline \nonumber
\mbox{net} : \ce{ CH4 + H2O} &\rightarrow \ce{CO + 3H2 } 
\end{split} 
\end{align}
\end{subequations}
More examples of pathways belonging to scheme (C) may be found in \cite{moses11,rm16,tsai17}.

\subsubsection{Comparison with previous work}

\begin{figure*}
\begin{center}
\vspace{-0.1in}
\includegraphics[width=\columnwidth]{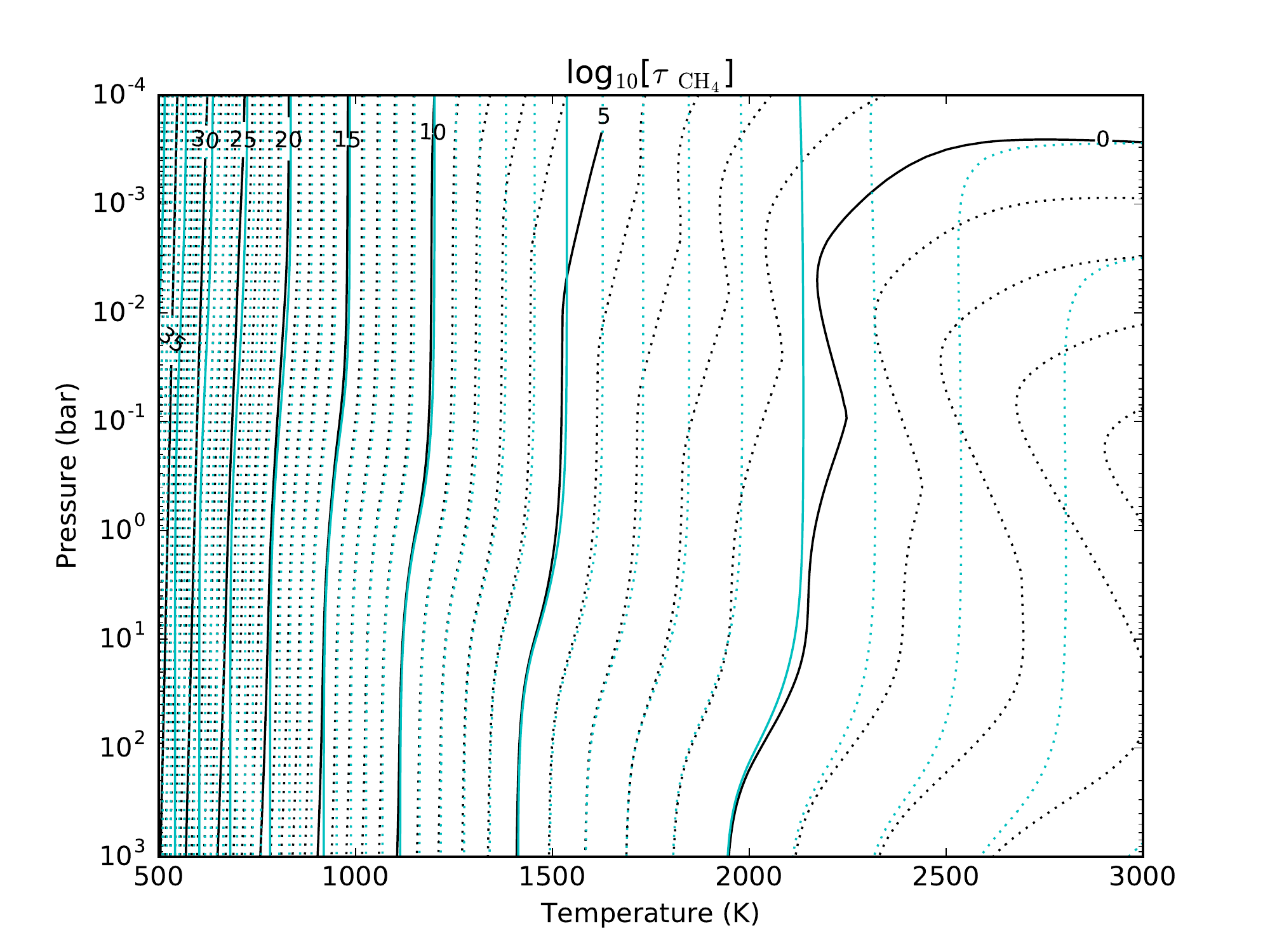}
\includegraphics[width=\columnwidth]{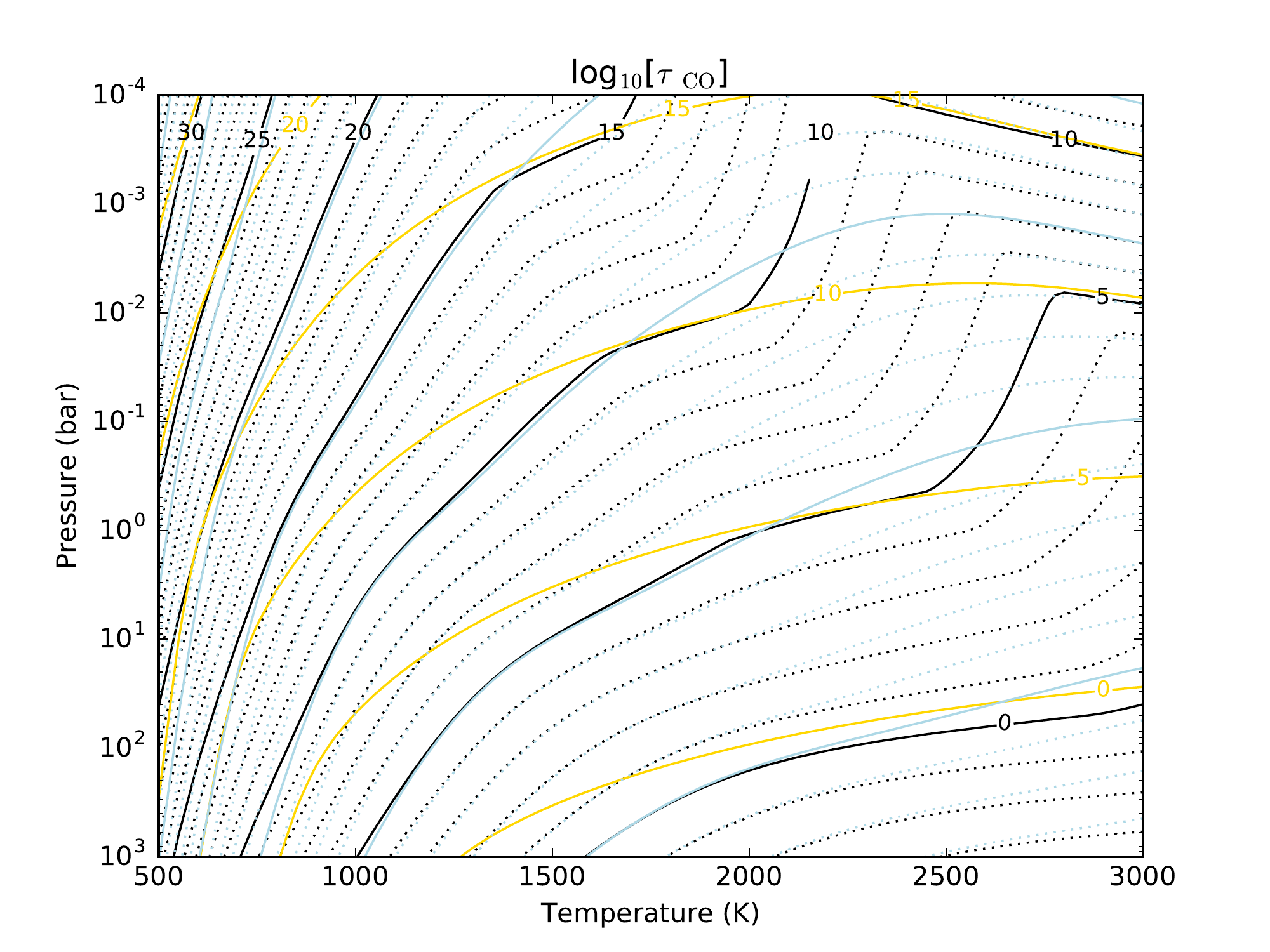}
\includegraphics[width=\columnwidth]{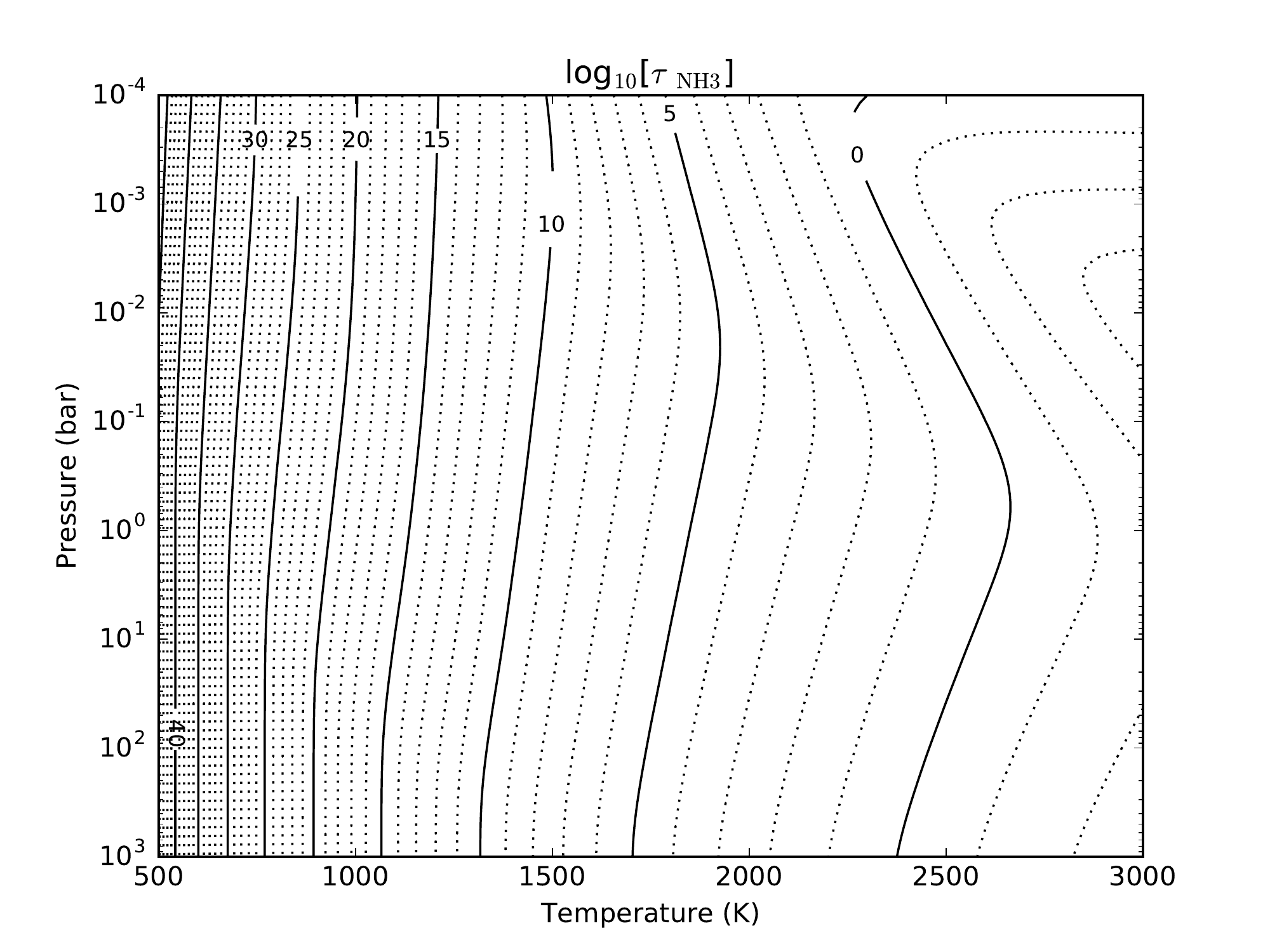}
\includegraphics[width=\columnwidth]{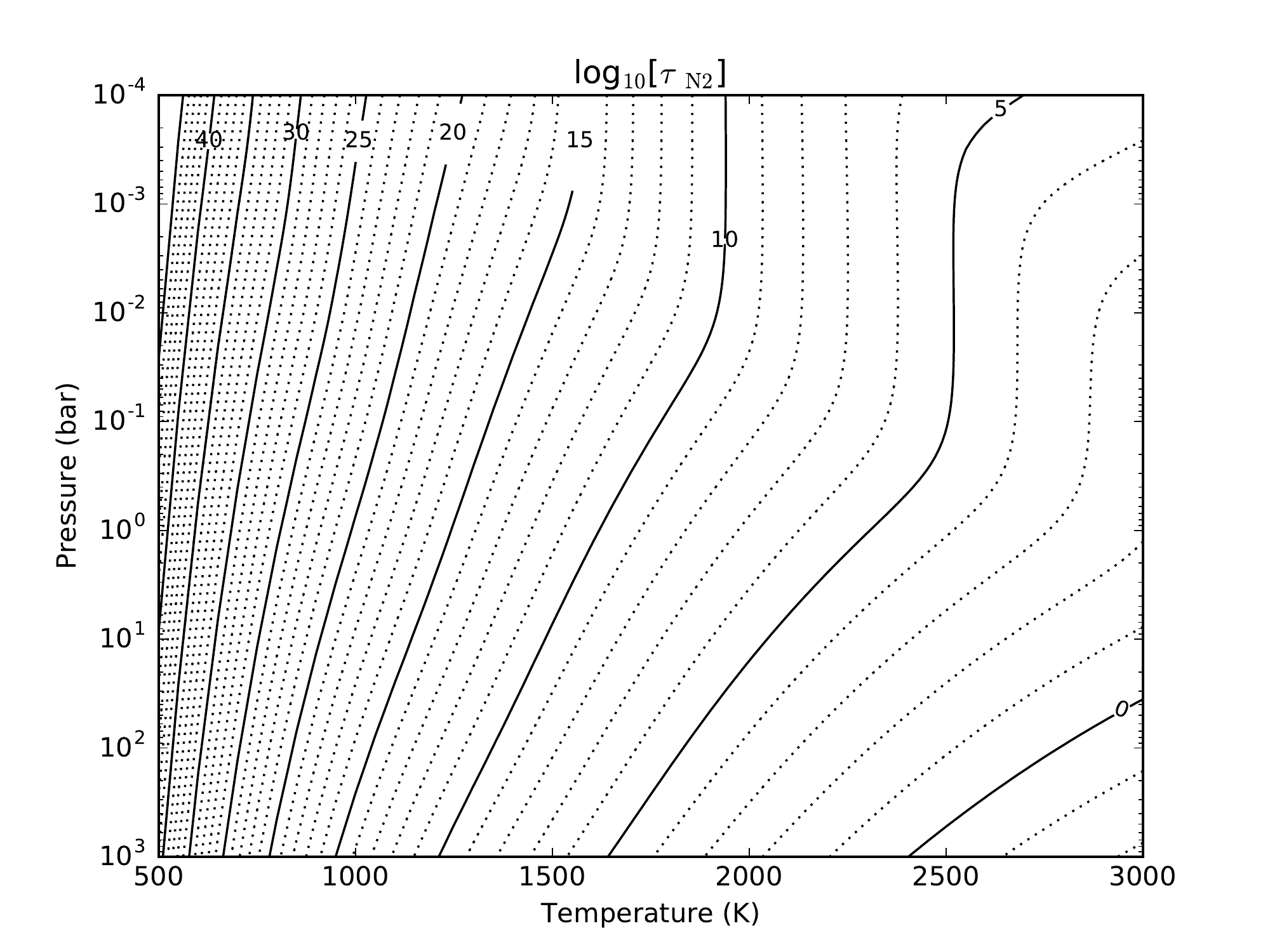}
\end{center}
\vspace{-0.2in}
\caption{Chemical timescales associated with the production of CH$_4$, CO, NH$_3$ and N$_2$ as functions of temperature and pressure.  The black curves are our calculations, while the cyan curves are from \cite{vc12} (for CH$_4$ and CO) and the yellow curves are from \cite{cs06} (for CO).}
\label{fig:tau_contours}
\end{figure*}

Previous work studying CO-CH$_4$ interconversion has typically assumed one or two RLS, because their intentions were to estimate the location of the quench point \citep{yung88,lf02,ms11,vm11}.  For the purpose of implementing chemical relaxation, this is insufficient.  In Figure \ref{fig:tau_contours}, we demonstrate this by comparing the calculations of chemical timescales associated with CH$_4$ and CO with those from \cite{cs06} and \cite{vc12}. The discrepancies between our calculations and those from \cite{vc12} are mainly present at high temperatures and low pressures. There are significant discrepancies between our calculations and the approximate ones by \cite{cs06} due to a different RLS assumed in \cite{cs06} as explained in the following paragraph. We use equation (16), instead of (19), of \cite{cs06}, because the latter is an approximation that is valid only when $T \lesssim 2000$ K and the mole fraction of H$_2$ is close to 1.  Instead of using equations (1) to (5) from \cite{cs06}, we perform the full chemical-equilibrium calculations.  We also include chemical timescales associated with NH$_3$ and N$_2$ for completeness.


\cite{cs06} consider a single CO-\ce{CH4} pathway
\begin{subequations}
\begin{align}
\ce{H2CO + H + M} &\rightarrow \ce{CH3O + M} \label{Ra}\\
\ce{CH3O + H2} &\rightarrow \ce{CH3OH + H} \label{Rb}\\
\ce{CH3OH + H} &\rightarrow \ce{CH3 + H2O} \label{Rc}
\end{align}
\end{subequations}
proposed by \cite{yung88}, where CO reacts with hydrogen to form \ce{H2CO} (formaldehyde)  and goes through \ce{CH3O} (methoxide) and \ce{CH3OH} (methanol) to get to \ce{CH3} (methyl).
They suggested (\ref{Ra}), which is involved in breaking the \ce{C=O} bond, as being the RLS.  \cite{cs06} adopt the rate constants from \cite{page} and \cite{bezard} for the low- and high-pressure limits, respectively.  We verify that the rate coefficient from \cite{page} has a value that is similar to what we use in our chemical network and is not the source of the discrepancies between our calculations and those of \cite{cs06}.   

Rather, the discrepancies stem from the rate coefficient associated with (\ref{Rb}), which had not been measured experimentally. \cite{yung88} assumed this reaction to be relatively fast, based on comparison with other similar reactions (see their Appendix A).  Motivated by the importance of CH$_3$OH kinetics (see \citealt{vc10} for details), \cite{moses11} performed ab initio calculations for the rate coefficients. According to their rate coefficients, (\ref{Rb}) always reacts slower than (\ref{Rc}) and thus should be the RLS \footnotemark[2] We have chosen to use the rate coefficients of \cite{moses11}. \footnotetext[2]{We also find this pathway suggested by \cite{yung88}, except for (\ref{Rb}) being the RLS instead of (\ref{Ra}), becomes dominant at high pressures ($\gtrsim$ 100 bar) and is important for Jupiter and Saturn where the \ce{CH4}-CO quench level is much deeper.} 

\cite{vc12} identified (\ref{R1}) and (\ref{R2}) as being the RLSs and adopt the rate coefficients from \cite{jasper}.  We have also taken the rate coefficients for (\ref{R1}) from \cite{jasper}, but the reverse rate coefficient for (\ref{R2}) from \cite{tsang}.  The differences between these rate coefficients are within a factor of 2. The two RLSs in \cite{vc12} control the most relevant temperature-pressure regions for hot/warm Jupiters ($\sim$ 1000 - 1500 K) and their timescale agrees well with our calculation until entering the high-temperature and low-pressure regime.

\subsection{NH$_3$-N$_2$ interconversion}

\begin{figure}[h]
\begin{center}
\vspace{-0.1in}
\includegraphics[width=0.9\columnwidth]{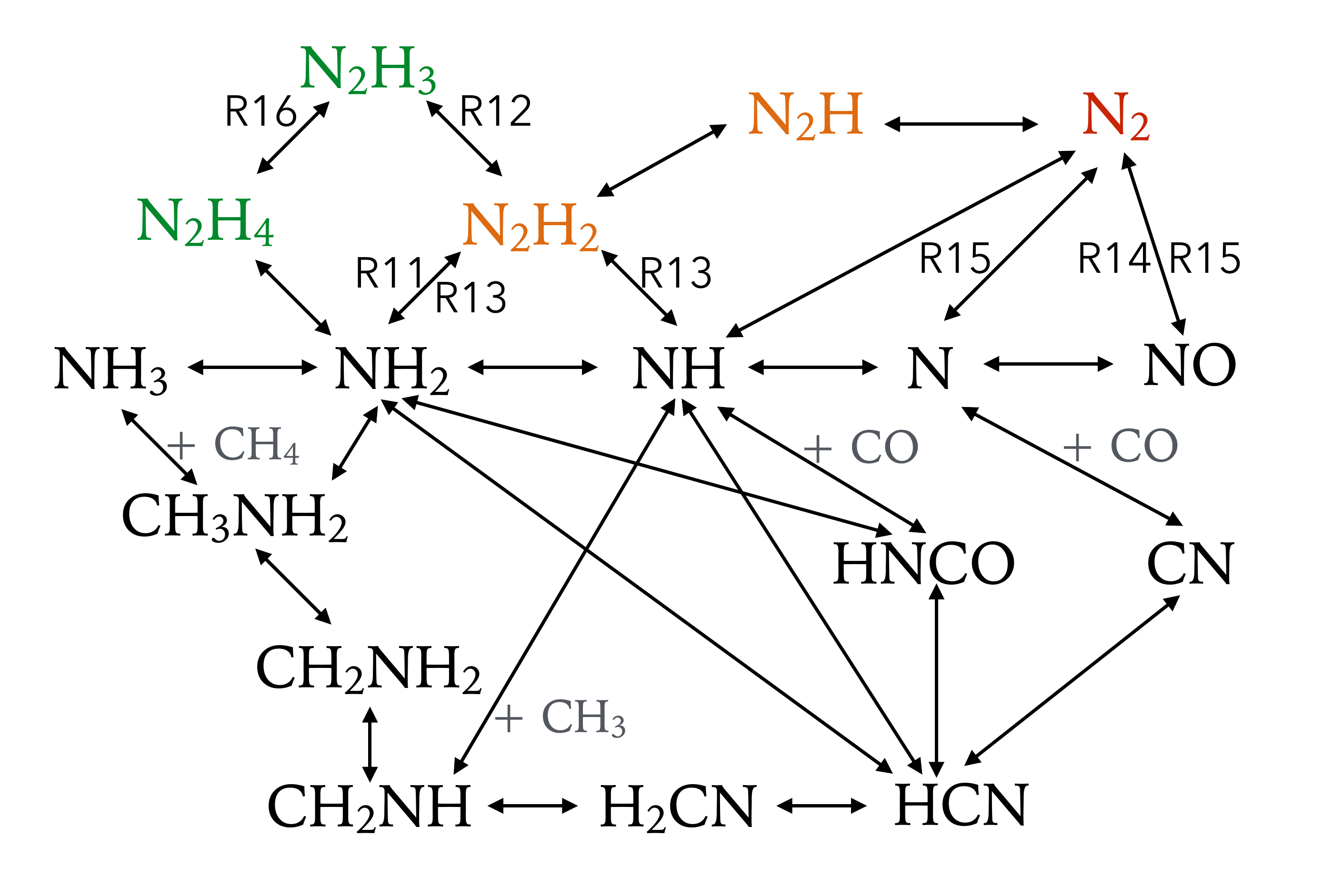}
\end{center}
\vspace{-0.2in}
\caption{Schematic illustration as Figure \ref{fig:scheme} but for the major chemical pathways between \ce{NH3}, \ce{N2}, and HCN.}
\label{fig:N_scheme}
\end{figure}

The timescales of nitrogen species are less constrained than that of CH$_4$-CO interconversion. HCN participates when $T$ $\gtrsim$ 1000 K and complicates the interconversion. We find it not straightforward to quantify the contribution of HCN and the real timescales deviate from those simply considering NH$_3$-N$_2$ interconversion.

The RLSs for NH$_3$-N$_2$ interconversion are
\begin{reactions}\label{R11}
\ce{NH2 + NH2 -> N2H2 + H2}
\end{reactions}

\begin{reactions}\label{R12}
\ce{N2H3 + M -> N2H2 + H + M}
\end{reactions}

\begin{reactions}\label{R13}
\ce{NH + NH2 -> N2H2 + H}
\end{reactions}

\begin{reactions}\label{R14}
\ce{NO + NH2 -> N2 + H2O}
\end{reactions}

\begin{reactions}\label{R15}
\ce{N + NO -> N2 + O}
\end{reactions}


\begin{reactions}\label{R16}
\ce{N2H4 + H -> N2H3 + H2}
\end{reactions}





Figure \ref{fig:N_scheme} visualizes the network of RLSs for nitrogen chemistry, while Table \ref{tab: pathways_N2} lists the RLSs of NH$_3$-N$_2$, across temperature and pressure. 

NH$_3$-N$_2$ interconversion can be divided into two schemes, depending on whether \ce{N2} is formed from \ce{N2H} or NO. At high pressures (the parameter space occupied by (\ref{R11}), (\ref{R12}), and (\ref{R13}) in Table \ref{tab: pathways_N2}), \ce{N2} is mainly formed by the dissociation of \ce{N2H}, with a pathway such as
\begin{subequations}
\label{re:N_highP}
\begin{align}
\begin{split} 
2 (\ce{ NH3 + H} &\rightarrow \ce{NH2 + H2 })\\
\ce{ NH2 + NH2} &\rightarrow \ce{N2H2 + H2 }\\ 
\ce{ N2H2 + H} &\rightarrow \ce{N2H + H2 }\\
\ce{ N2H + M} &\rightarrow \ce{N2 + H + M }\\
\ce{ H2 + M }&\rightarrow \ce{ 2H + M }\\
\hline \nonumber
\mbox{net} : \ce{ 2NH3} &\rightarrow \ce{N2 + 3H2 } 
\end{split} 
\end{align}
\end{subequations}
where the second reaction (\ref{R11}) is the RLS. As temperature increases, (\ref{R11}) is replaced by (\ref{R13}), or a channel through \ce{N2H3} and \ce{N2H4}  (\ref{R12}). This pathway is similar to that identified for HD 209458b by \cite{moses11}.

At low pressures (the parameter space occupied by (\ref{R14}) and (\ref{R15}) in Table \ref{tab: pathways_N2}), H$_2$ is attacked by the more abundant free atomic O and produces OH, which in turn forms NO with N via \ce{ N + OH } $\rightarrow$ \ce{ NO + H }. NO then react with N or NH$_2$, depending on the temperature, to produce N$_2$. This step involves forming the N$\equiv$N bond and is usually the RLS. The pathway becomes 
\begin{subequations}
\label{re:N_highP}
\begin{align}
\begin{split} 
2(\ce{ NH3 + H} &\rightarrow \ce{NH2 + H2 })\\
2(\ce{ NH2 + H} &\rightarrow \ce{NH + H2 })\\
2(\ce{ NH + H} &\rightarrow \ce{N + H2 })\\
\ce{ N + NO }&\rightarrow \ce{ N2 + O }\\
\ce{ O + H2 }&\rightarrow \ce{ OH + H }\\
\ce{ N + OH }&\rightarrow \ce{ NO + H }\\
2(\ce{ H2 + M }&\rightarrow \ce{ 2H + M })\\
\hline \nonumber
\mbox{net} : \ce{ 2NH3} &\rightarrow \ce{N2 + 3H2 } 
\end{split} 
\end{align}
\end{subequations}
where (\ref{R15}) is the RLS.

\begin{table} [!h]
\begin{center}
\caption{\ce{NH3} $\leftrightarrow$ \ce{N2} rate-limiting reactions}
\label{tab: pathways_N2}
\begin{tabular}{cccccc}
\hline
\textbf{T (K)} & 500 &  1000  &  1500  &  2000  &  2500  \\
\hline
\textbf{P (bar)} &&&&&\\
10$^{-4}$ & R11 & R14 & R15 & R15 & R15\\
10$^{-3}$ & R11 & R14 & R13 & R15 & R15\\
10$^{-2}$ & R11 & R13 & R13 & R15 & R15\\
10$^{-1}$ & R11 & R12 & R13 & R13 & R15\\
1         & R12 & R12 & R13 & R13 & R13\\
10        & R11 & R12 & R13 & R13 & R13\\
100       & R11 & R12 & R16 & R13 & R13\\
\hline
\end{tabular}
\end{center}
\end{table}

\section{Computing chemical timescales}
\label{sec:chemtime}

The full expressions for the chemical timescales, including those involving methane, carbon monoxide, water, ammonia and molecular nitrogen, are stated in Appendix \ref{app:f_tau}.  In the following, we explain the reasoning behind their construction.

\subsection{Revisiting CH$_4$-CO interconversion}

\begin{table}
\begin{center}
\caption{\ce{CH4} $\leftrightarrow$ \ce{CO} rate-limiting reactions (solar abundance)}
\label{tab: pathways}
\begin{tabular}{cccccc}
\hline
\textbf{T (K)} & 500 &  1000  &  1500  &  2000  &  2500  \\
\hline
\textbf{P (bar)} &&&&&\\
10$^{-4}$ & R9 & R2 & R3 & R5 & R6 \\
10$^{-3}$ & R9 & R2 & R3 & R4 & R5 \\
10$^{-2}$ & R9 & R2 & R2 & R4 & R5 \\
10$^{-1}$ & R9 & R1 & R2 & R3 & R4 \\
1         & R9 & R1 & R2 & R3 & R3 \\
10        & R9 & R1 & R1 & R3 & R3 \\
100       & R9 & R1 & R1 & R7 & R7 \\

\hline
\end{tabular}
\end{center}
\end{table}

\begin{table}
\begin{center}
\caption{\ce{CH4} $\leftrightarrow$ \ce{CO} rate-limiting reactions (C/O = 2)}
\label{tab: pathways_CtoO2}
\begin{tabular}{cccccc}
\hline
\textbf{T (K)} & 500 &  1000  &  1500  &  2000  &  2500  \\
\hline
\textbf{P (bar)} &&&&&\\
10$^{-4}$ & R9 & R2 & R8  & R8 & R5 \\
10$^{-3}$ & R9 & R2 & R8  & R8 & R8 \\
10$^{-2}$ & R9 & R2 & R10 & R8 & R8 \\
10$^{-1}$ & R9 & R1 & R2  & R8 & R8 \\
1         & R9 & R1 & R2  & R3 & R8 \\
10        & R9 & R1 & R1  & R3 & R8 \\
100       & R9 & R1 & R7  & R7 & R7 \\
\hline
\end{tabular}
\end{center}
\end{table}

The study of interconversion between methane and carbon monoxide has a long and rich history. In the current study, we focus on identifying the pathways in reducing atmospheres, while the same steps can be applied to other types of atmospheres. In this subsection, our goal is to provide an analytical expression for computing the timescale associated with CH$_4$-CO interconversion as a function of the reactions (\ref{R1}) to (\ref{R10}).  Table \ref{tab: pathways} shows the RLSs for solar metallicity, while Table \ref{tab: pathways_CtoO2} shows them for $\mbox{C/O}=2$.  Reactions (\ref{R8}) and (\ref{R10}) are only relevant for $\mbox{C/O}=2$, as previously mentioned.

\begin{figure}
\begin{center}
\vspace{-0.1in}
\includegraphics[width=0.75\columnwidth]{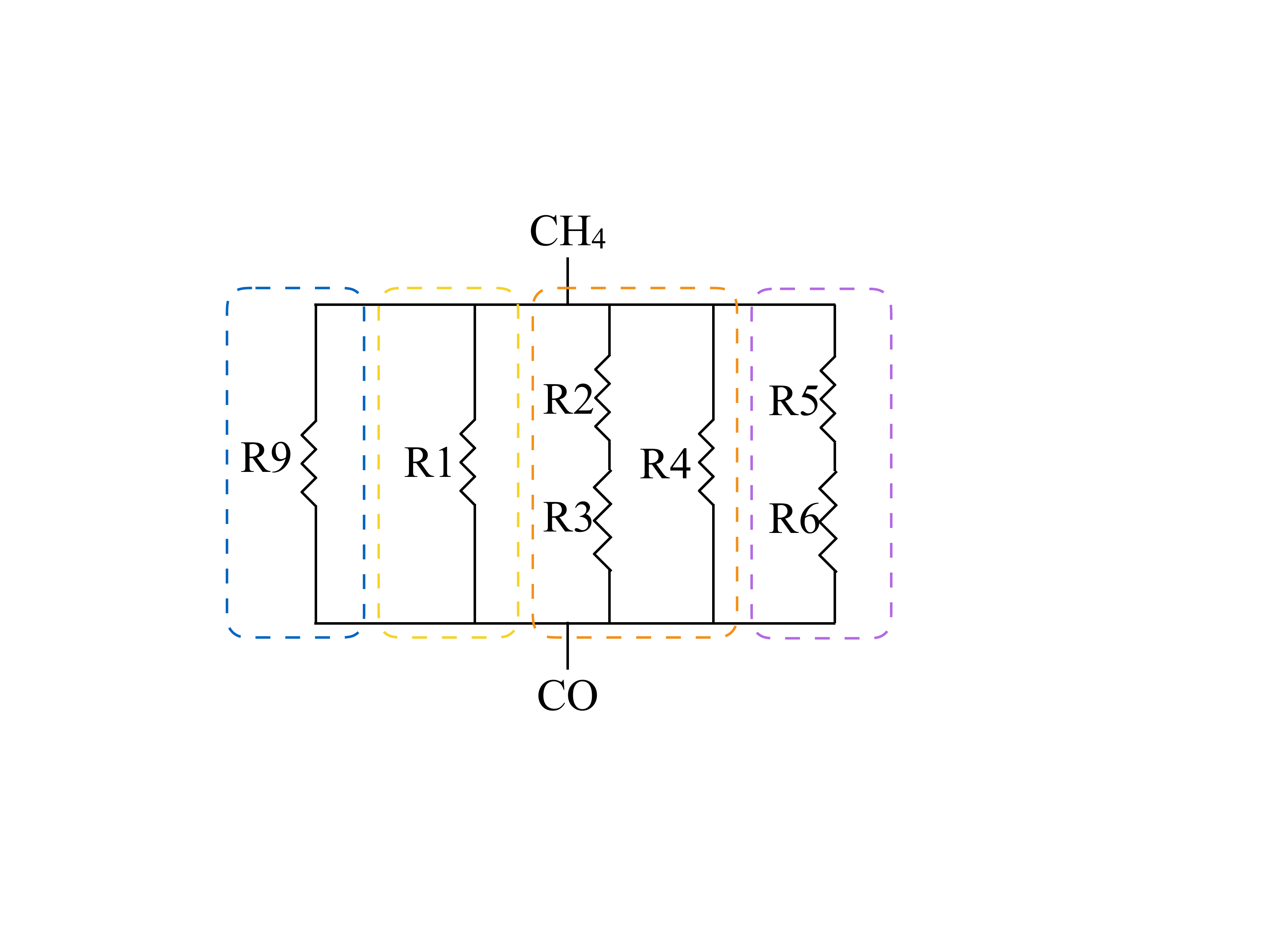}
\hspace*{-0.3cm}\includegraphics[width=0.67\columnwidth]{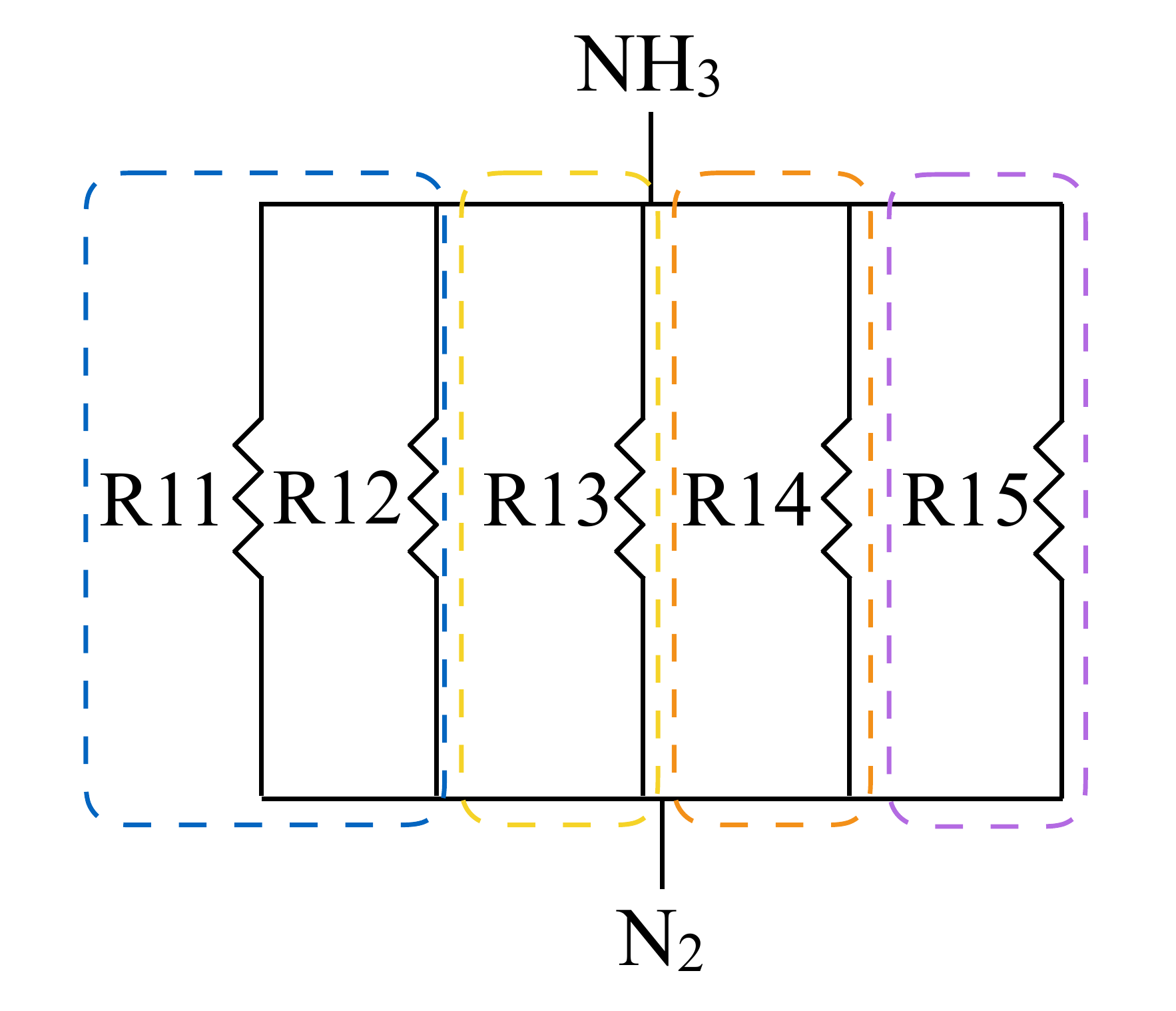}
\end{center}
\vspace{-0.2in}
\caption{The effective ``electronic circuit" of the major chemical pathways between methane and carbon monoxide (upper) for Figure \ref{fig:scheme} and between ammonia and molecular nitrogen (lower) for Figure \ref{fig:N_scheme}. The dashed rectangles in blue, yellow, orange, and purple group the RLSs according to their operating temperatures and pressures.}
\label{fig:circuit}
\end{figure}

Under differing conditions of temperature and pressure, the various RLSs can either collaborate or compete with one another.  We find that a useful analogy for understanding the pathways is to visualize them as the resistors in an electrical circuit---built in series or in parallel.  Using such an analogy, we can construct an analytical expression for the chemical timescale that consists of a network of reactions working either in series or in parallel. The upper diagram of Figure \ref{fig:circuit} visualizes how such an analogous electrical circuit would look like for CH$_4$-CO interconversion. We identify the series and parallel RLSs and group the RLSs that operate in similar temperatures and pressures together. For example, (\ref{R2}), (\ref{R3}) are in series but (\ref{R2}) and (\ref{R4}) are in parallel.  Depending on the temperature and pressure, either (\ref{R1}), (\ref{R9}), or the group of (\ref{R2}), (\ref{R3}), and (\ref{R4}) is in control.

Mathematically, a pair of reactions in parallel can be expressed as an operation that takes the maximum of the two rate coefficients.  For a pair of reactions in series, the operation instead takes the minimum of the two rate coefficients. The groups that operate in particular temperatures and pressures are then added together for simplicity. In this way, the series of relationships between the reactions can be expressed as 
\begin{equation}
\begin{split}
\tau_{\rm CH_4} &= \frac{[\mbox{CH}_4]}{\mbox{r1} + \max{\left( \min{\left( \mbox{r2,r3} \right)}, \mbox{r4} \right)} + \mbox{r9} + \min{\left( \mbox{r5,r6} \right)} }\\ 
& + \tau_{\rm H_2}\times \frac{3 [{\rm CO}]}{[{\rm H_2}]},
\end{split}
\end{equation}
where [X] represents the equilibrium number density of species X in cm$^{-3}$, rx stands for the reaction rate of (Rx), and the factor $\frac{3 [{\rm CO}]}{[{\rm H_2}]}$ stems from the amount of hydrogen that participates in the \ce{CH4}-CO interconversion (three \ce{H2} for every CO according to the net reaction). The second term is usually orders of magnitude smaller than the first term except at high temperatures and low pressures, where the dissociation and recombination between H and H$_2$ become important and the relatively slower conversion of hydrogen starts to bottleneck the process. We demonstrate how methane is controlled by hydrogen in Figure \ref{fig:sensitivity} where we manually vary the rate of H-H$_2$ dissociation/recombination.

We have also replaced (\ref{R7}) with (\ref{R1}) for simplicity because the rates of both reactions are very similar at high pressures. The same formula is applied to $\tau_{\rm CO}$ except one needs to replace the numerator with [CO], since the interconversion goes both ways.  

\subsection{Water}

H$_2$O reacts efficiently with atomic hydrogen via 
\begin{reactions}\label{R18}
\ce{H2O + H -> OH + H2}
\end{reactions}
in a hydrogen-rich atmosphere. Being the major oxygen carrier, water prevailingly participates in various reaction pathways, e.g. CO $\leftrightarrow$ CO$_2$ and H$_2$ $\leftrightarrow$ 2H. For solar metallicity, carbon is the richest heavy element next to oxygen. We find the timescale of water is effectively determined by the interconversion rate of CH$_4$-CO,
\begin{equation} \label{eq:tau_f}
\begin{split}
\tau_{\ce{H2O}} &= \frac{[\mbox{H}_2\mbox{O}]}{\mbox{r1} + \max{\left( \min{\left( \mbox{r2,r3} \right)}, \mbox{r4} \right)} + \mbox{r9} + \min{\left( \mbox{r5,r6} \right)} }\\
& + \tau_{\rm H_2}\times \frac{3 [{\rm CO}]}{[{\rm H_2}]}.
\end{split}
\end{equation}
\subsection{Carbon dioxide}

Carbon dioxide is produced through the relatively fast scheme \citep{lly10,moses11} 
\begin{equation*}
\begin{split} 
\ce{H2O + H} & \rightarrow \ce{ OH + H2 }\\
\ce{CO + OH} & \rightarrow \ce{ CO2 +H }\\
\hline \nonumber
\mbox{net} : \ce{CO + H2O} & \rightarrow \ce{CO2 + H2} 
\end{split} 
\end{equation*}
The chemical timescale is simply
\begin{equation}\label{eq:tau_co2} 
\tau_{\mbox{CO}_2} = \frac{[\mbox{CO}_2]}{k_{\mbox{CO}_2}[\mbox{CO}][\mbox{OH}]}
\end{equation}
where $k_{\mbox{CO}_2}$ is the rate coefficient for \ce{CO + OH -> CO2 +H }.  Due to the fast conversion, CO$_2$ still maintains pseudo-equilibrium with CO and H$_2$O after the latter two are quenched, before CO$_2$ reaches its own quench point (see the discussion in section 3.1 of \citealt{moses11}). Owing to this coupling, instead of relaxing CO$_2$ to its equilibrium value, we find it correct to relax CO$_2$ toward the pseudo-equilibrium value determined by the (possibly quenched) CO and H$_2$O, as expressed in equation (\ref{eq:co2_pseudo}), similar to the treatment in equation (43) of \cite{zm14}.

\subsection{Ammonia and molecular nitrogen}

The chemical timescale of ammonia is approximately determined by the \ce{NH3}-\ce{N2} conversion. Omitting HCN, we group the RLSs with similar operating temperatures and pressures and construct the effective electrical circuit in the lower diagram of Figure \ref{fig:circuit}. 

For ammonia production, we have 
\begin{equation}
\tau_{\rm NH_3} = \frac{1}{2} \left( \frac{[\mbox{NH}_3]}{ \max{\left(\mbox{r11,r12}\right)} + \mbox{r13} + \mbox{r14}   + \mbox{r15}} +\tau_{\rm H_2}\times \frac{3 [{\rm N2}]}{[{\rm H_2}] } \right)
\end{equation}
where the factor $\frac{3 [{\rm N2}]}{[{\rm H_2}]}$ is again the amount of hydrogen that participates in the \ce{NH3}-\ce{N2} interconversion, as in the \ce{CH4}-CO interconversion limited by H-H$_2$ interconversion at high temperatures and low pressures, and the factor of 1/2 comes from the fact that the net reaction converts two NH$_3$ molecules to one N$_2$ molecule.  

Following the same steps, the chemical timescale of molecular nitrogen is expressed as
\begin{equation}
\tau_{\rm N2} = \frac{[\mbox{N}_2]}{ \max{\left(\mbox{r11,r12}\right)} + \mbox{r13} + \mbox{r14}   + \mbox{r15}} +\tau_{\rm H_2}\times \frac{3 [{\rm N2}]}{[{\rm H_2}] }
\end{equation}

\subsection{The effects of metallicity}

As the metallicity is varied from $10^{-2} \times$ to $1000 \times$ of the solar values, we find that the chemical pathways and RLSs discussed in section \ref{sec:rls} do not change. Therefore, the formulae for the timescales remain the same. An exception is for the NH$_3$-N$_2$ pathways as the metallicity approaches $100 \times$ solar. (\ref{R14}) and (\ref{R15}) occupy more of the temperature-pressure parameter space because of the richness of oxygen.  We also confirm that $\tau_{\rm CO}$ is almost independent of metallicity and $\tau_{\rm CH_4}$ is inversely proportional to metallicity, as found by \cite{vc12}.  

Once the metallicity increases beyond $1000 \times$ solar, the atmosphere ceases to be H$_2$-dominated and the main constituents become carbon dioxide or molecular oxygen \citep{hu14}.  For example, at $10^4\times$ solar metallicity, CH$_4$ becomes scarce simply because of the lack of hydrogen for its formation.  In this scenario, CO-CO$_2$ interconversion becomes the main quenching process and follows the same pathway as for solar metallicity with the timescale still given by equation (\ref{eq:tau_co2}).  The pathways involving the nitrogen species become completely altered.  Generally, the reactions of NH$_{\rm x}$ with H$_2$O become important in controlling NH$_3$-N$_2$ interconversion.

\subsection{The effects of C/O}

The carbon-to-oxygen ratio (C/O) is a crucial factor in controlling the atmospheric chemistry and thermal structure \citep{madhu12,moses13a,venot15,rocc16}. Equilibrium chemistry is sensitive to C/O and undergoes a qualitative transition at $\mbox{C/O} = 1$. We explore C/O values ranging from 0.1 to 2. The reason to limit ourselves at $\mbox{C/O} = 2$ is that C/O much larger than unity is considered unlikely as the surplus carbon tends to condense and form graphite \citep{moses13b}. Comparing Table \ref{tab: pathways} and Table \ref{tab: pathways_CtoO2}, the major change as C/O exceeds unity is carbon takes the route through C$_2$H$_2$ at high temperature.  The typical pathway in a hot, carbon-rich atmosphere is 
\begin{subequations}
\begin{align}
\begin{split} 
\ce{ CH4 + H} &\rightarrow \ce{ CH3 + H2 } \\
\ce{ CH3 + CH2} &\rightarrow \ce{ C2H4 + H} \\
\ce{ C2H4 + H} &\rightarrow \ce{ C2H3 + H2} \\
\ce{ C2H3 + M} &\rightarrow \ce{ C2H2 + H + M} \\
\ce{ C2H2 + O} &\rightarrow \ce{ CH2 + CO} \\
\ce{ OH + H} &\rightarrow \ce{ O + H2 } \\
\ce{ H + H2O }&\rightarrow \ce{ OH + H2 } \\
\ce{ H2 + M}&\rightarrow \ce{ 2H + M } \\
\hline \nonumber
\mbox{net} : \ce{ CH4 + H2O} &\rightarrow \ce{CO + 3H2 } 
\end{split} 
\end{align}
\end{subequations}
where \ce{ C2H2 + O} $\rightarrow$ \ce{CH2 + CO } (\ref{R8}) is the RLS in the above pathway. The corresponding timescale for C/O between 1 and 2 is
\begin{equation}
\tau_{\rm CH_4} = \frac{[\mbox{CH}_4]}{\mbox{r1} + \min{\left( \mbox{r2,r3} \right)} + \mbox{r9} + \max{\left( \mbox{r8,r10} \right)}},
\end{equation}
In this scheme, CH$_4$-CO conversion is no longer limited by hydrogen dissociation.

\section{Comparison of chemical networks using pathway analysis}
\label{sec:pathway}

\begin{figure}
\begin{center}
\vspace{-0.1in}
\includegraphics[width= 0.3\columnwidth]{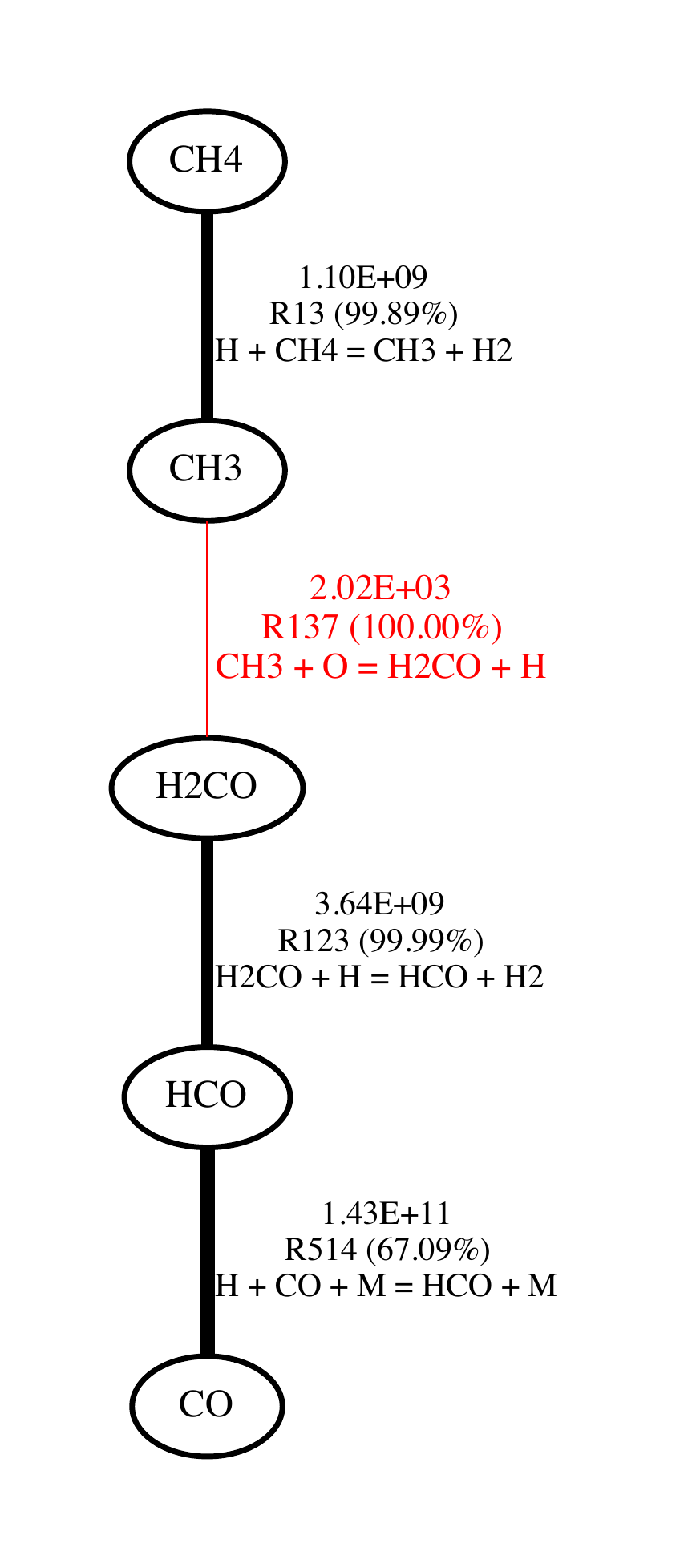}
\includegraphics[width= 0.3\columnwidth]{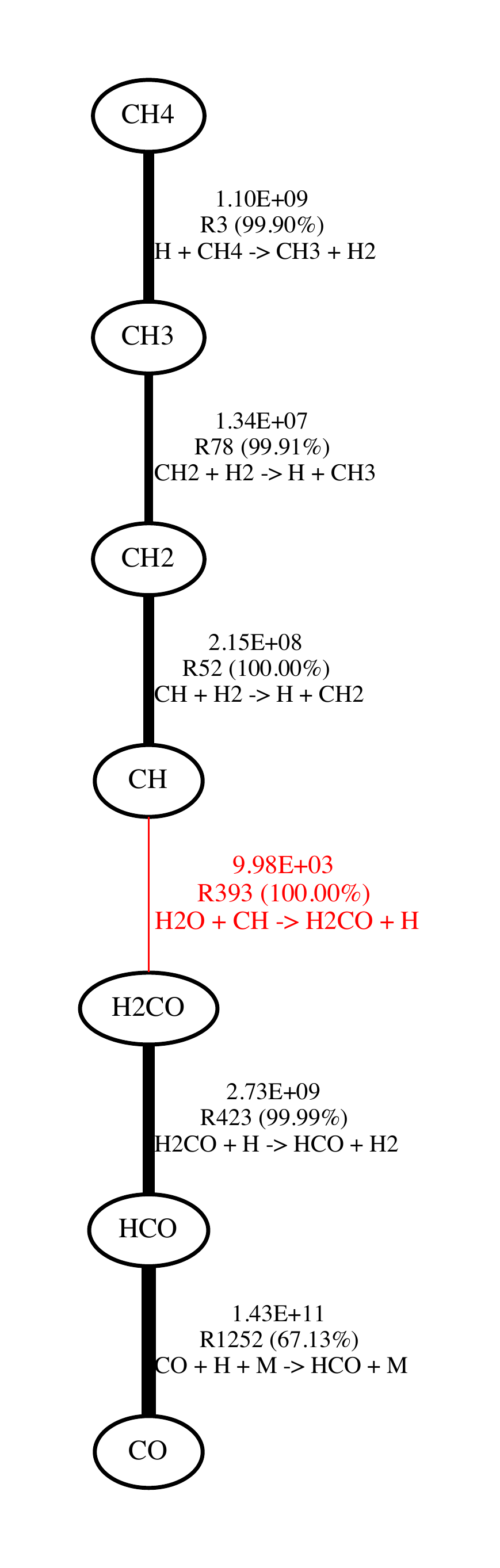}
\includegraphics[width= 0.3\columnwidth]{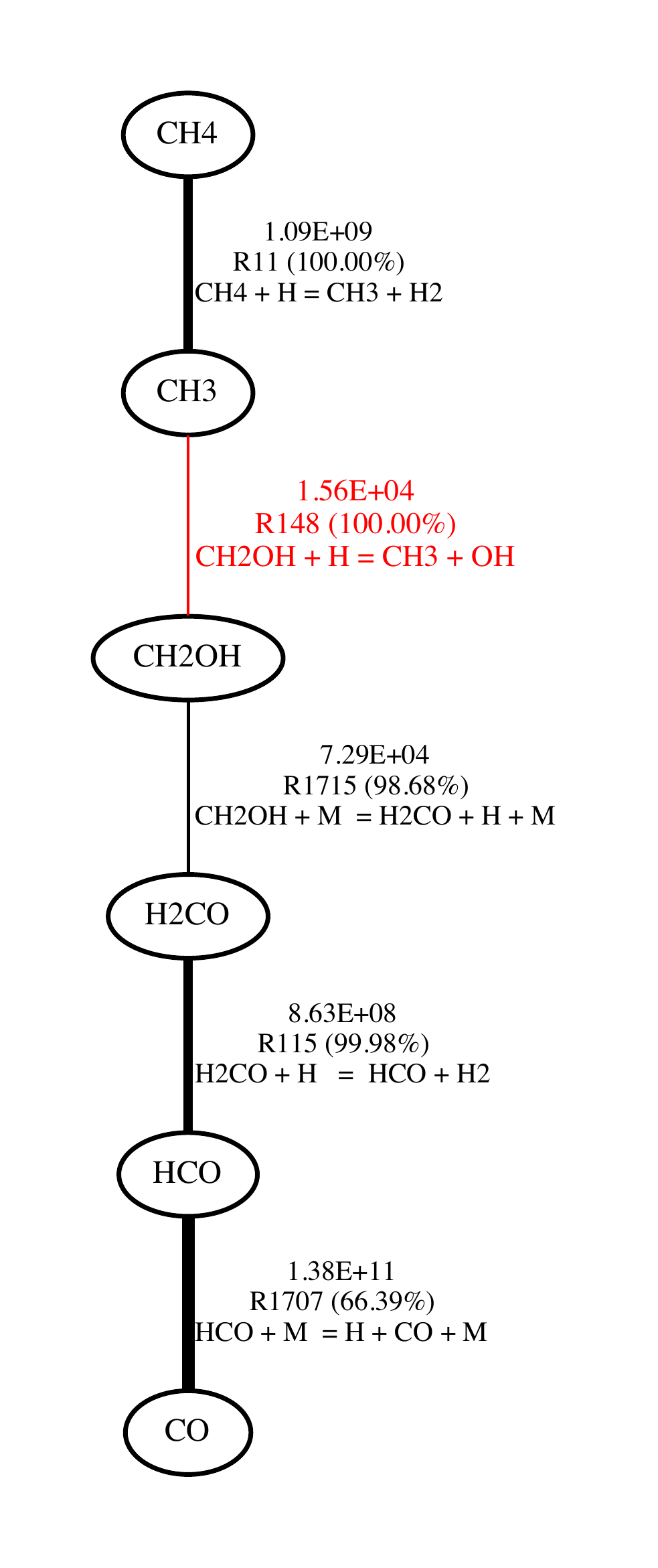}
\end{center}
\vspace{-0.2in}
\caption{Examples of CH$_4$-CO pathway analysis at $T = 2000$ K and $P = 0.1$ bar with the chemical network from VULCAN (left), \cite{moses11} (middle), and \cite{venot12} (right). Thicker lines represent faster reaction rates (denoted by the first-row numbers shown in in cm$^{-3}$s$^{-1}$ and the percentage of contribution to the interconversion rate is also provided) and the red lines are the rate-limiting steps.}
\label{fig:path_tool}
\end{figure}

We recognize that the chemical pathways taken depend entirely on the network one is using. With a different network, the rate-limiting chemical reactions need to be re-identified. The expressions for chemical timescales can be worked out following the same steps in \S\ref{sec:chemtime}. Here, we compare our network with two others that have included high-temperature chemical kinetics and been applied to hot exoplanets.  We first compare our network with that of \cite{moses11}; our two networks are naturally similar, because we have taken the values of certain rate coefficients from that study.  We then compare our network with that of \cite{venot12}.  In Figure \ref{fig:path_tool}, we use our pathway analysis tool to compare the route taken by each network (Since the reaction rates are calculated with equilibrium composition, the forward rate equals to the reverse rate, there is no directionality in the pathways, i.e. CH$_4$ $\rightarrow$ CO takes the same pathway as CO $\rightarrow$ CH$_4$). 

At solar abundance and temperatures less than 2000 K, the network of \cite{moses11} shows almost exactly the same CH$_4$-CO pathways as ours. Due to a different choice of the rate coefficients associated with (\ref{R2}), part of the parameter space occupied by (\ref{R3}) is replaced by (\ref{R2}) in their network for $T \gtrsim 2000$ K and $P \sim 10$ bar.  (\ref{R7}) also extends to lower pressures in their network.  At high temperatures and low pressures in scheme (A), their network experiences the same dehydrogenation process, but instead of (\ref{R4}) and (\ref{R5}), their network chooses a faster path through water: \ce{CH + H2O} $\rightarrow$ {H2CO + H} (which is not included in our network).  Yet, in this regime, the timescale is limited by hydrogen dissociation and yields the same timescale of CH$_4$-CO interconversion as we do.  At $\mbox{C/O} = 2$, carbon forms abundant \ce{C2H2} and then gets oxidized to CO in a similar way, but except via (\ref{R10}), it takes \ce{C2H2 + OH} $\rightarrow$ \ce{H2CCO + H} or \ce{C2H2 + O} $\rightarrow$ {HCCO + H} in their network. \ce{H2CCO} or \ce{HCCO} then proceeds to be split into CO by H. In general, our network is consistent with that of \cite{moses11}, as suggested by the comparison in \cite{tsai17}. 

The overall CH$_4$--CO timescale in \cite{venot12} is shorter than ours and \cite{moses11}.  We find the two key reactions that make the RLSs and essentially the timescales different are (\ref{R9}) and 
\begin{reactions}\label{V1}
\ce{CH2OH + M} \rightarrow \ce{H2CO + H + M}.
\end{reactions} 
\cite{venot12} includes faster rate coefficients for both (\ref{R9}) and (\ref{V1}).  Their rate coefficient for (\ref{R9}) is based on the work of \cite{hidaka}, which has been suggested as overestimating the rate; see discussion in \cite{vc12} and \cite{moses14}. This rate coefficient is significantly higher ($\sim 10$ orders of magnitudes) than the ab initio calculation in \cite{moses11}, which is also used in our network. 

At lower temperatures (T $\lesssim$ 1000 K), \cite{venot12} takes the same pathway through (\ref{R9}), forming \ce{CH3OH} from \ce{CH3} and \ce{H2O}. However, with a faster rate it never bottlenecks the pathway and controls the timescale. Their RLSs are instead the reactions involving forming or destroying \ce{H2CO}, e.g. \ce{CH3O + M} $\rightarrow$ \ce{H2CO + H + M} and \ce{H2CO + H} $\rightarrow$ \ce{HCO + H2}. Similarly, due to the faster \ce{CH3}-\ce{CH3OH} channel via (\ref{R9}), for $1000 \mbox{ K} \lesssim T \lesssim 1500$ K, \cite{venot12} exhibits pathways close to ours, except that our (\ref{R1}) is replaced by \ce{CH2OH + H2} $\rightarrow$ \ce{CH3OH + H} as the RLS. At higher temperature where $T \gtrsim 1500$ K, the differences are mainly attributed to (\ref{V1}). \cite{venot12} uses the rate from \cite{gh86}, validated for 600--1000 K, while this work and \cite{moses11} use the rate from \cite{cribb92}, validated for 1900--2700 K. The former is about two orders of magnitude larger than the latter in this temperature range. Consquently, the fast \ce{CH2OH}-\ce{H2CO} interconversion in \cite{venot12} again never limits the pathway (e.g. the right pathway in Figure \ref{fig:path_tool}). Their RLS remains \ce{CH2OH + H2} $\rightarrow$ \ce{CH3OH + H} for high pressures and switches to \ce{CH3 + OH} $\rightarrow$ \ce{CH2OH + H} for low pressure. 

In conclusion, our pathway analysis tool is useful in identifying the key reactions for a given network, which allows us to diagnose the divergent behaviors of different networks.  By isolating the rate coefficients of the key reactions involved, we hope to motivate future laboratory and/or theoretical studies that will hopefully resolve these discrepancies.

\section{Validation of chemical relaxation method}
\label{sec:validate}

\begin{figure}
\begin{center}
\vspace{-0.1in}
\includegraphics[width=\columnwidth]{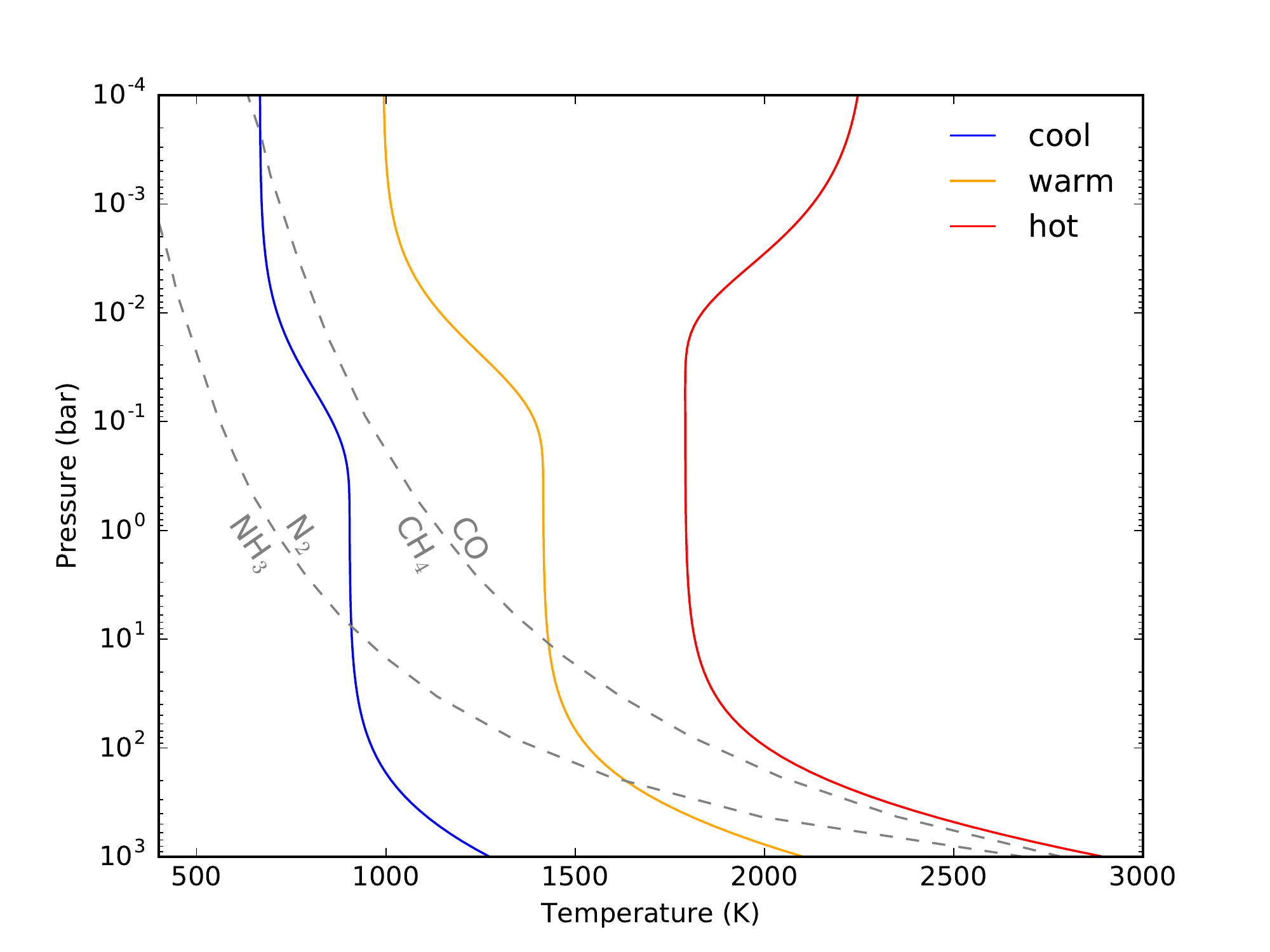}
\end{center}
\vspace{-0.2in}
\caption{Representative temperature-pressure profiles representing hot, warm, and cool atmospheres. Gray dashed curves show the boundaries where \ce{CH4}-CO and \ce{NH3}-\ce{N2} have equal abundances in chemical equilibrium.}
\label{fig:KTP}
\end{figure} 

We are finally ready to validate the chemical relaxation method to a factor of several, having assembled the necessary ingredients.
\subsection{Setup}

The goal of any chemical calculation, either from full kinetics or any simplified method, is to provide the rate of change of every species locally. The applicability of the relaxation method should not depend on the format or complexity of atmospheric dynamics. Therefore, before including the chemical relaxation method in a GCM, we evaluate whether and to what extend the relaxation method can replace full chemical kinetics with a one-dimensional full kinetics model, where eddy diffusion ($K_{zz}$) is used to represent vertical mixing.  For $w$ (vertical velocity) $\sim 1$ km s$^{-1}$ and $H \sim 100$ km, we have $K_{zz} \lesssim w H \sim 10^{12}$ cm$^2$ s$^{-1}$.  To exaggerate the effects of vertical mixing, we explore a range of $K_{zz}$ values up to $10^{15}$ cm$^2$ s$^{-1}$.

We run the same atmospheric conditions with the relaxation method and \texttt{VULCAN}, a full kinetics model with a C-H-O network including 29 species with up to two carbon atoms and about 300 forward and reverse reactions \citep{tsai17}. For nitrogen chemistry, an updated N-C-H-O network is implemented including 53 species and about 600 forward and reverse reactions. The chemical equilibrium abundances are calculated using the \texttt{FastChem}\citep{fastchem} code. We perform our chemistry calculations over three temperature-pressure profiles constructed using the analytical formula in \cite{hml14} to represent cool, warm and hot atmospheres (Figure \ref{fig:KTP}). These profiles are meant to mimic GJ 1214b-, HD 189733b- and WASP-18b-like atmospheres. With the atmospheres we tested, the computational time for integrating one step using the chemical relaxation method is about 5 ms, compared to about 0.5 s using \texttt{VULCAN}. The computational speed can be increased by about 100 times with the chemical relaxation method.

\subsection{Hot atmospheres (WASP-18b-like)}

\begin{figure*}
\begin{center}
\vspace{-0.1in}
\includegraphics[width= 0.87\columnwidth]{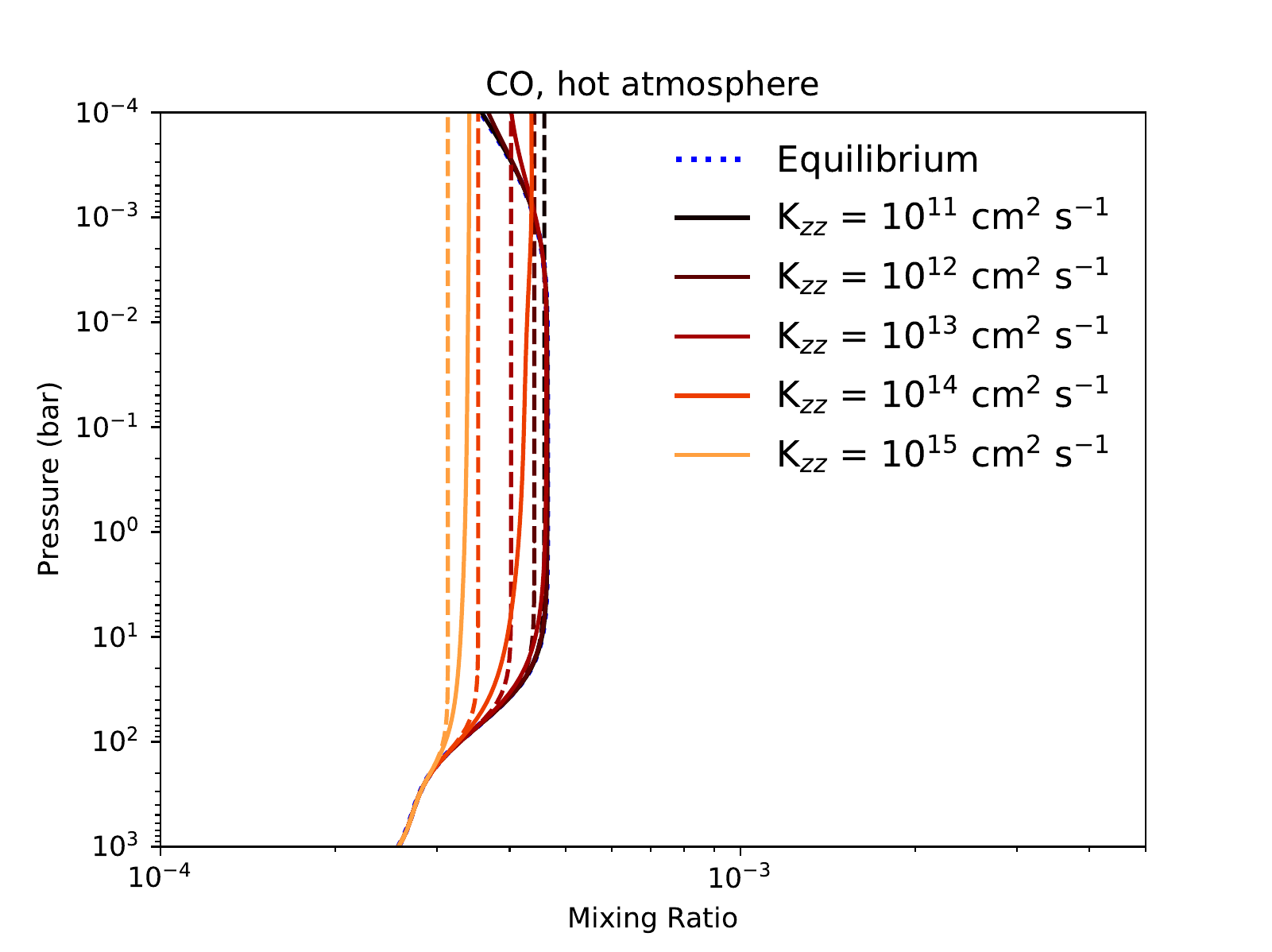}
\includegraphics[width= 0.87\columnwidth]{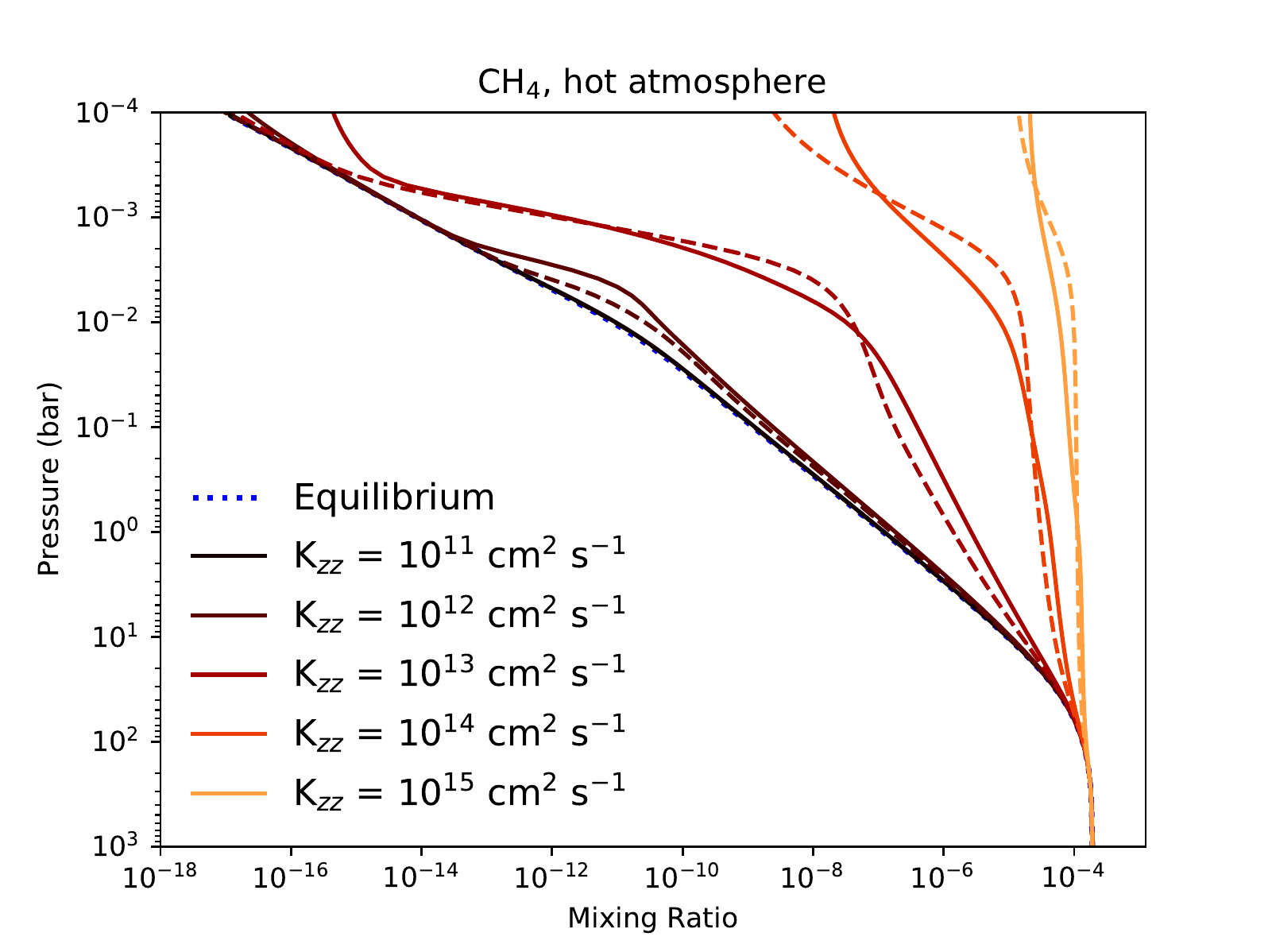}
\includegraphics[width= 0.87\columnwidth]{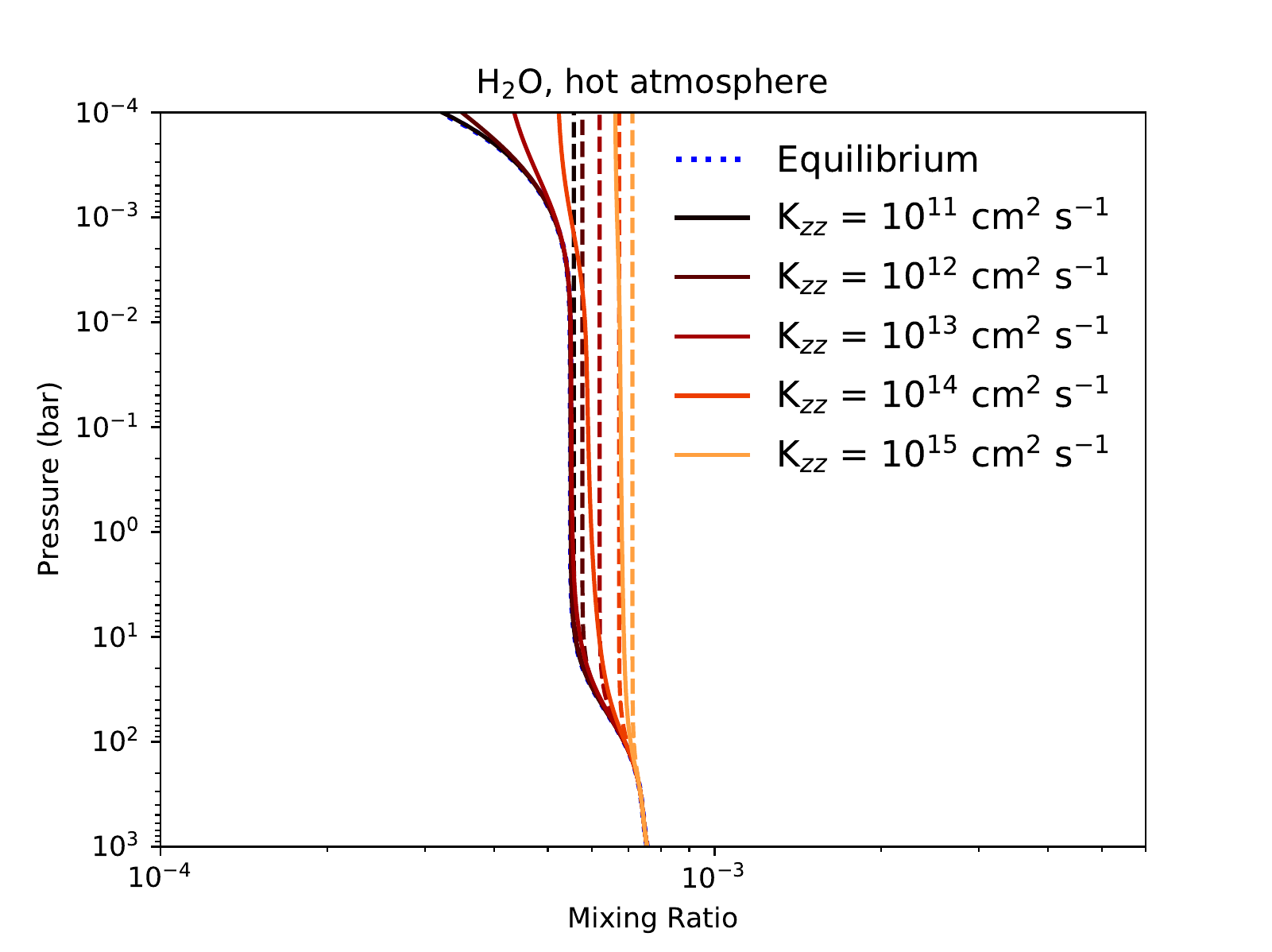}
\includegraphics[width= 0.87\columnwidth]{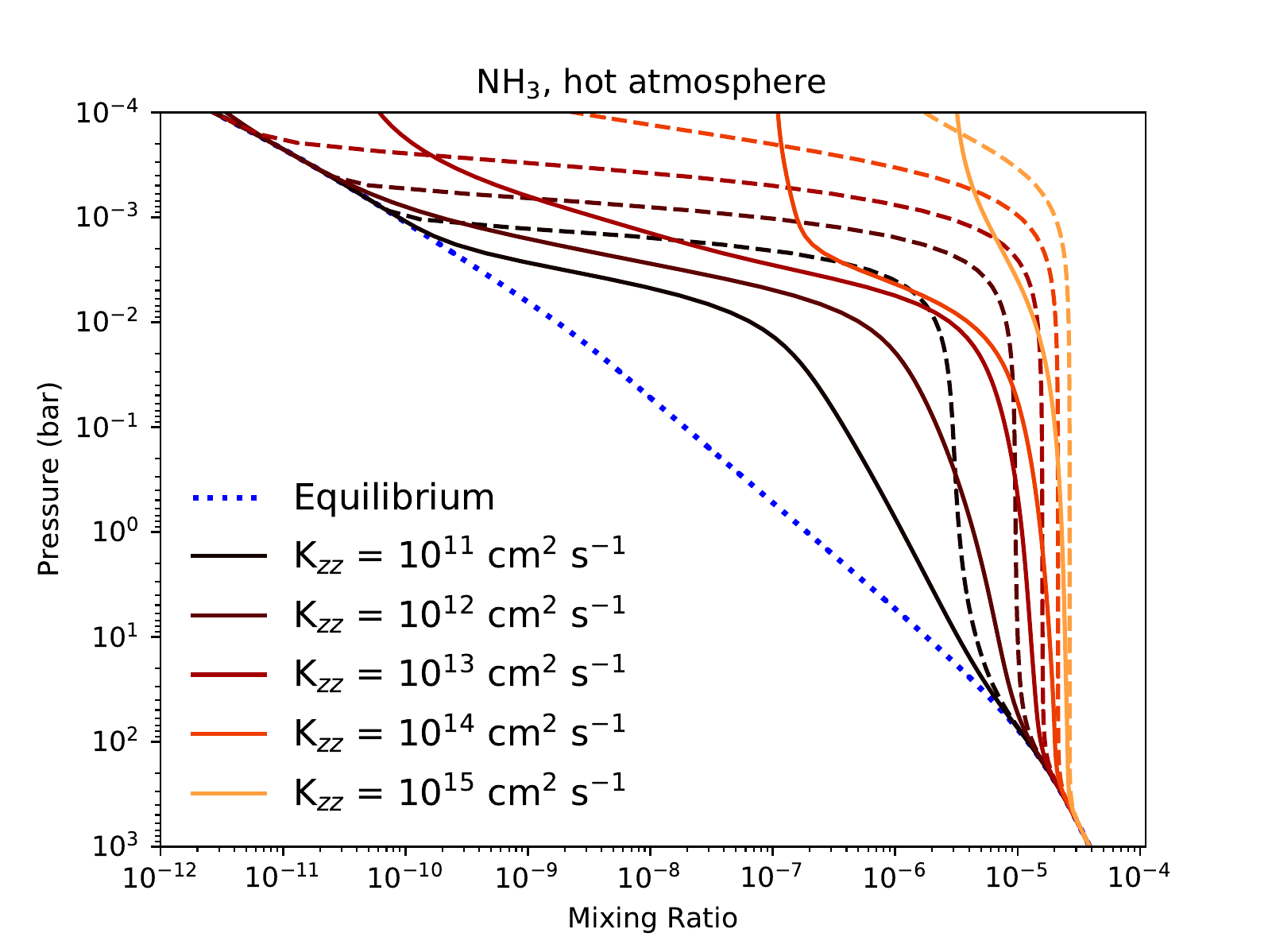}
\end{center}
\caption{Mixing ratios of CO,CH$_4$, H$_2$O, and NH$_3$ in the hot atmosphere as displayed in Figure \ref{fig:KTP}. The results of the relaxation method (dashed) are compared to the full chemical kinetics (solid) for a range of vertical mixing strengths shown in various colors.  The mixing ratio in chemical equilibrium is shown as a dotted curve.}
\label{fig:hot_valid}
\end{figure*}

At solar metallicity and high temperatures, CO, H$_2$O, and N$_2$ are the dominant molecules in chemical equilibrium \citep{madhu12,moses11,ht16}.  In Figure \ref{fig:hot_valid}, it is therefore unsurprising that the mixing ratios of the first two molecules (not showing N$_2$) are insensitive, or nearly independent of, pressure.  For $K_{zz} \lesssim 10^{13}$ cm$^2$ s$^{-1}$, the mixing ratios of CO and H$_2$O essentially track their chemical-equilibrium values closely, deviating only with stronger mixing and/or lower pressures ($\lesssim 1$ mbar).

The mixing ratios of CH$_4$ and NH$_3$ exhibit a much larger range of values as they both drop off significantly with increasing altitude: 13 and 7 orders of magnitude for CH$_4$ and NH$_3$, respectively, over the range of pressure examined (0.1 mbar to 1 kbar).  Over these broad ranges, the chemical relaxation method performs fairly well, exhibiting an accuracy of within an order of magnitude for the most part. With strong vertical mixing, CH$_4$ and NH$_3$ begin to quench at about 10 bar and 100 bar, respectively. At about 1 mbar, hydrogen dissociation/recombination slows down the interconversion and sets the second quench level. If the effect of hydrogen dissociation is neglected, the estimated chemical timescale will be too short and the prediction of CH$_4$ and NH$_3$ will be too close to chemical equilibrium.

\subsection{Warm atmospheres (HD 189733b-like)}

\begin{figure*}[!h]
\begin{center}
\vspace{-0.1in}
\includegraphics[width= 0.87\columnwidth]{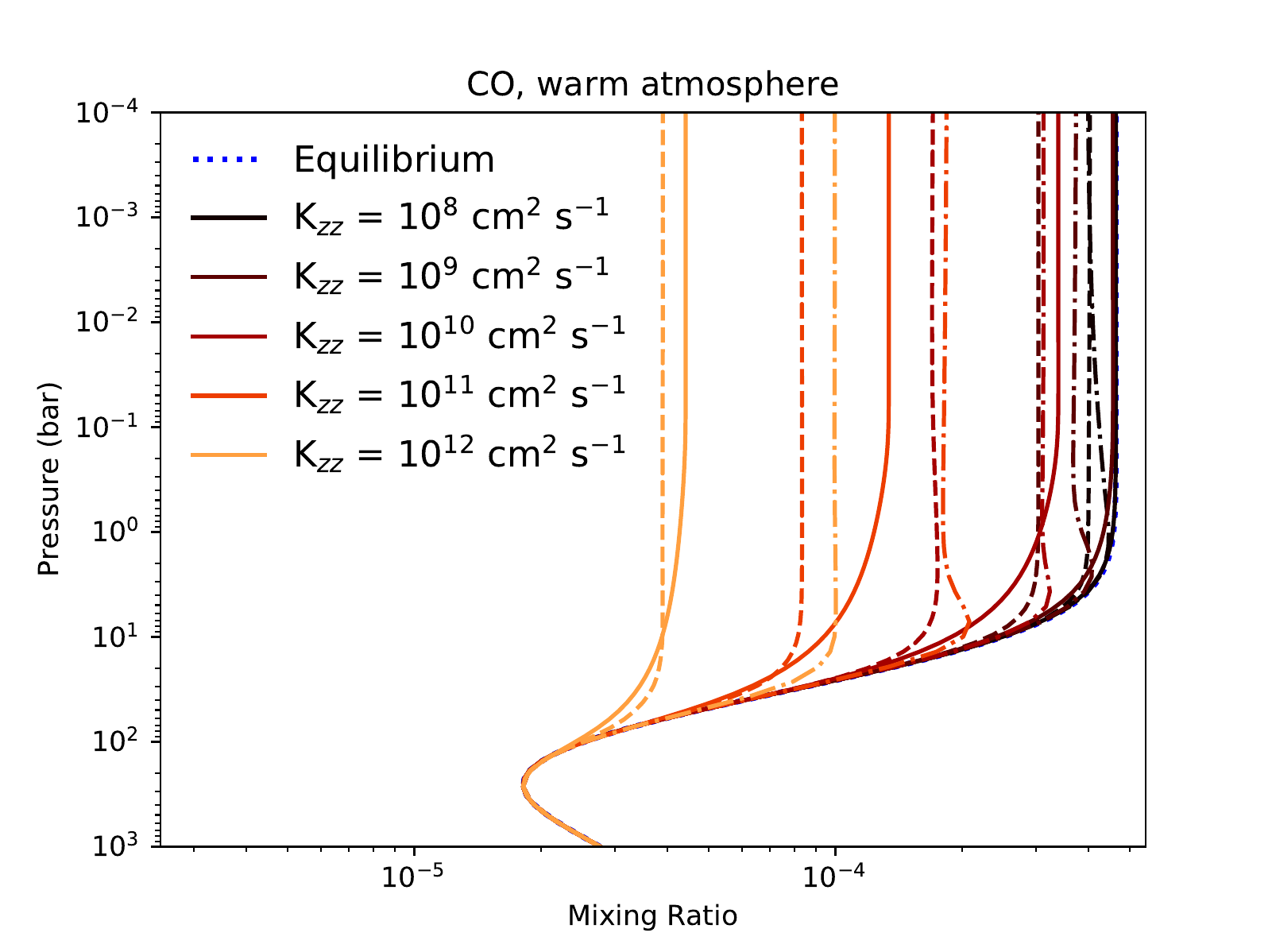}
\includegraphics[width= 0.87\columnwidth]{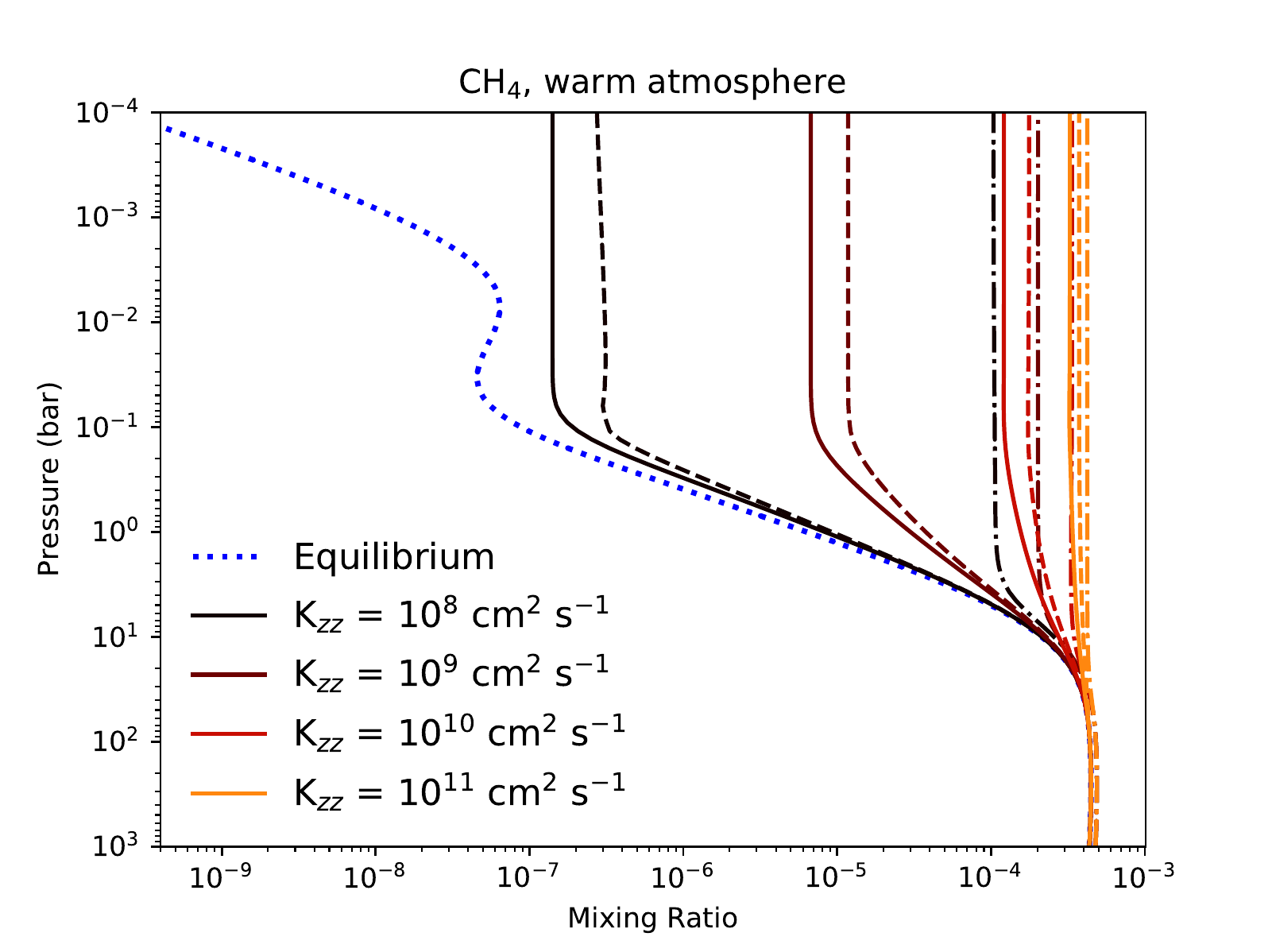}
\includegraphics[width= 0.87\columnwidth]{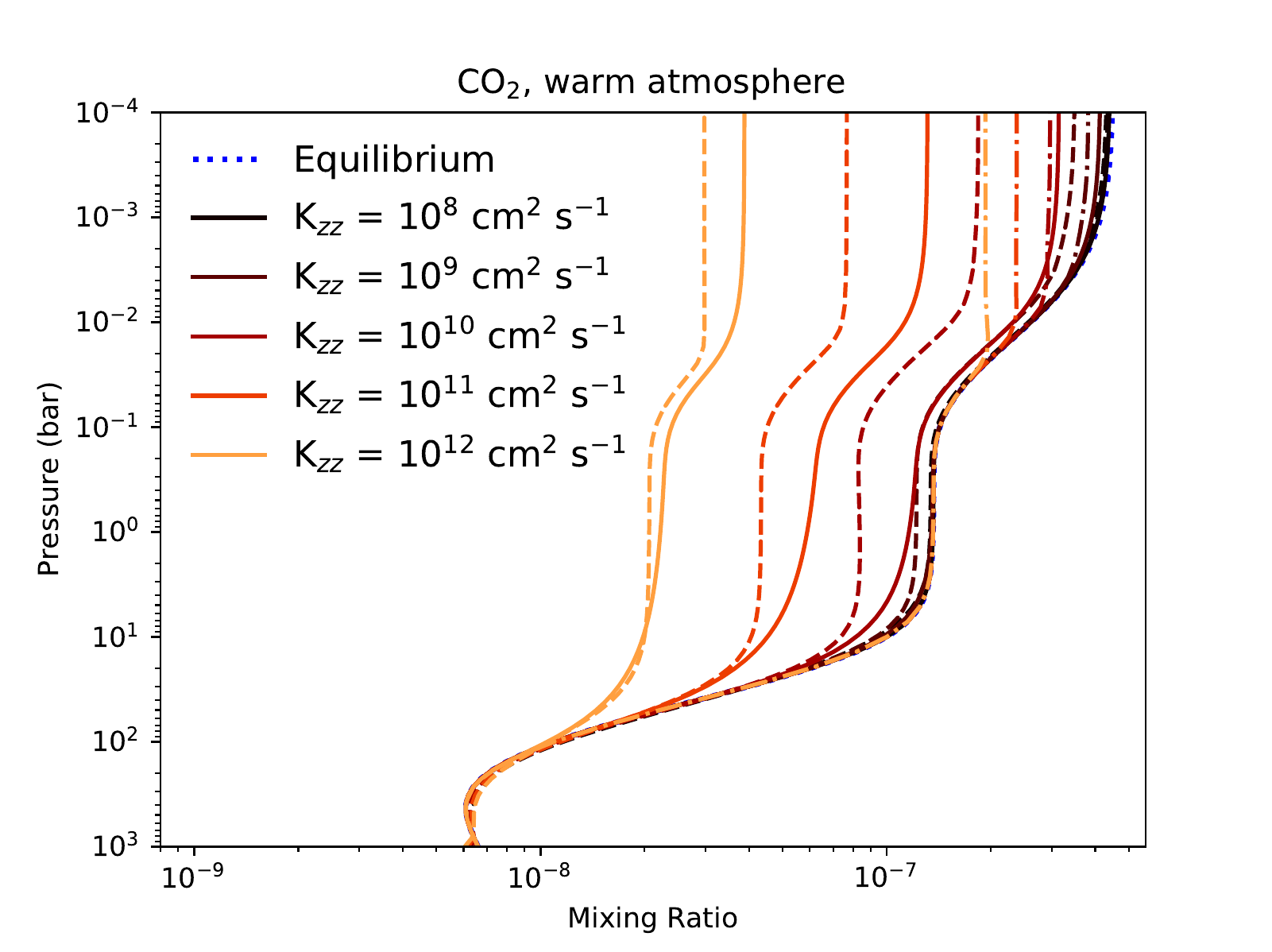}
\includegraphics[width= 0.87\columnwidth]{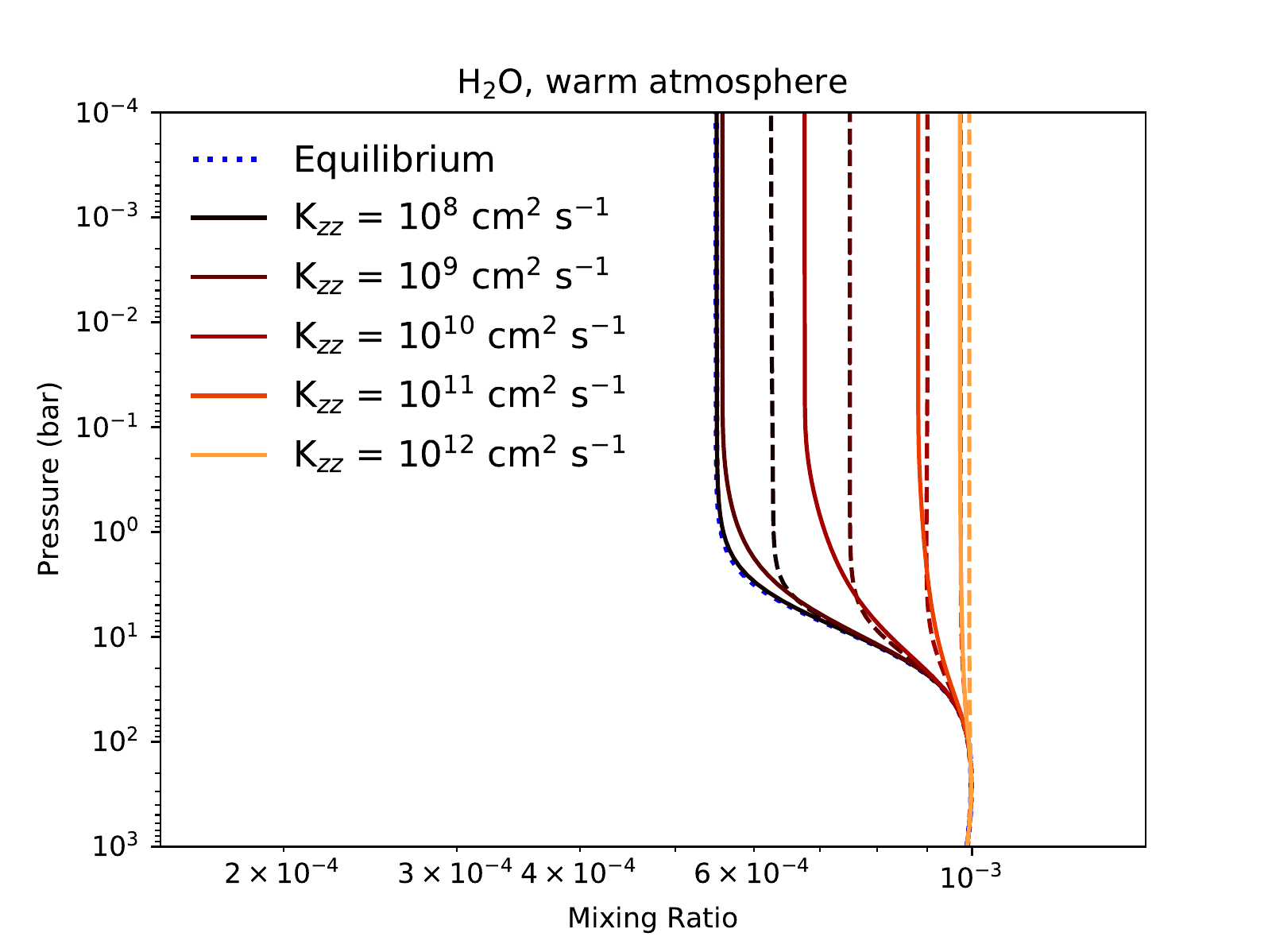}
\includegraphics[width= 0.87\columnwidth]{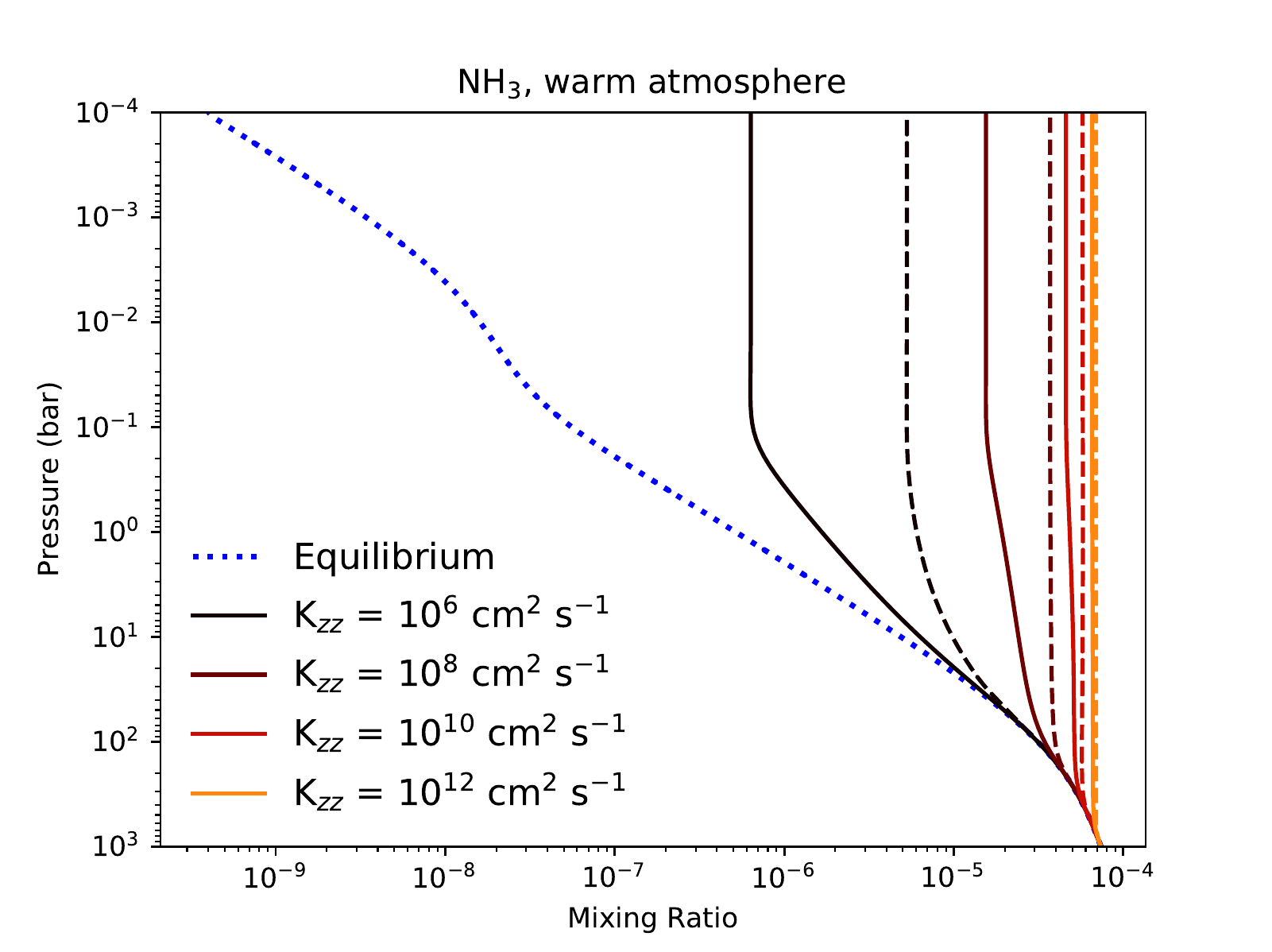}
\includegraphics[width= 0.87\columnwidth ]{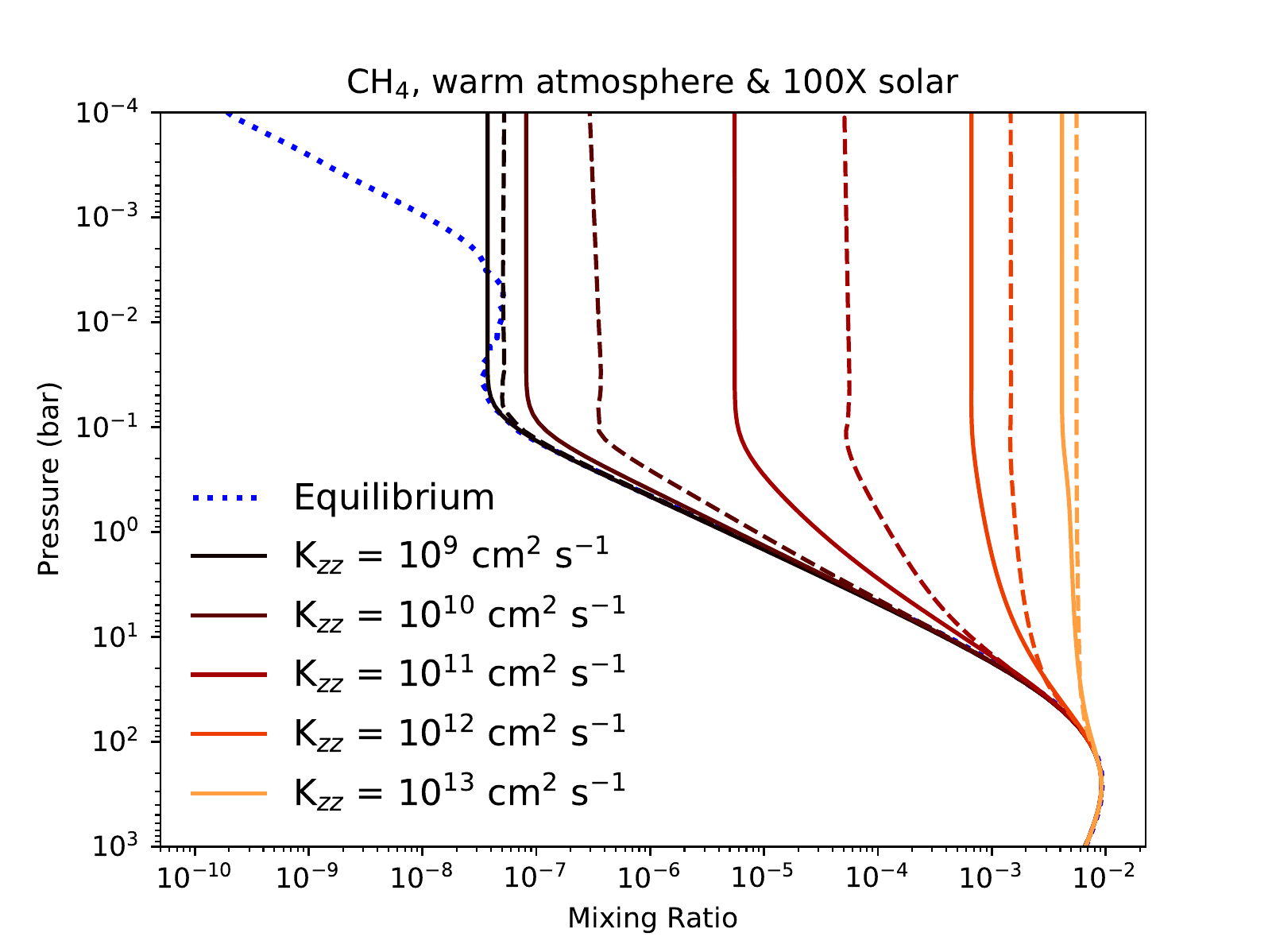}
\includegraphics[width= 0.87\columnwidth ]{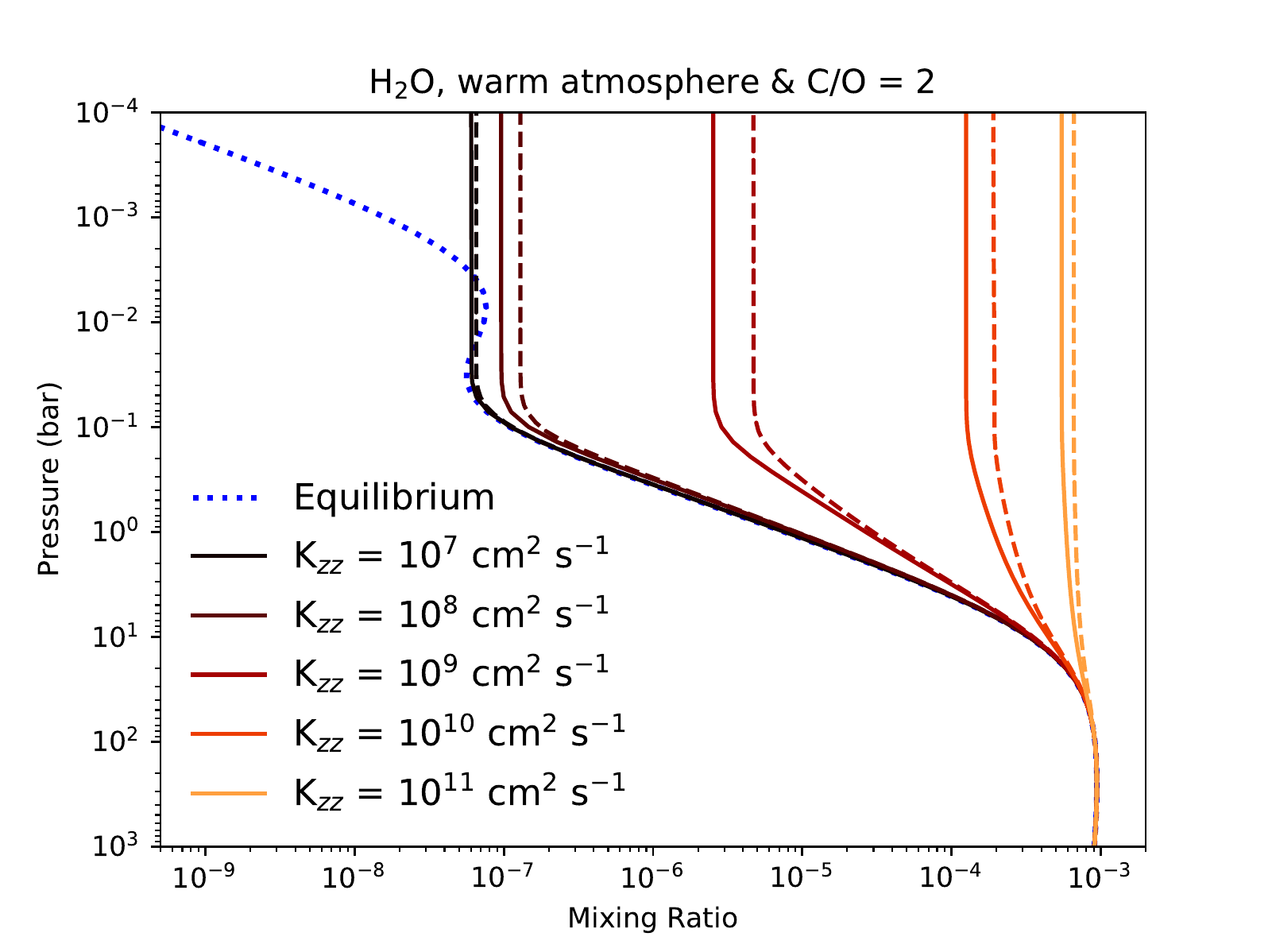}
\includegraphics[width= 0.87\columnwidth ]{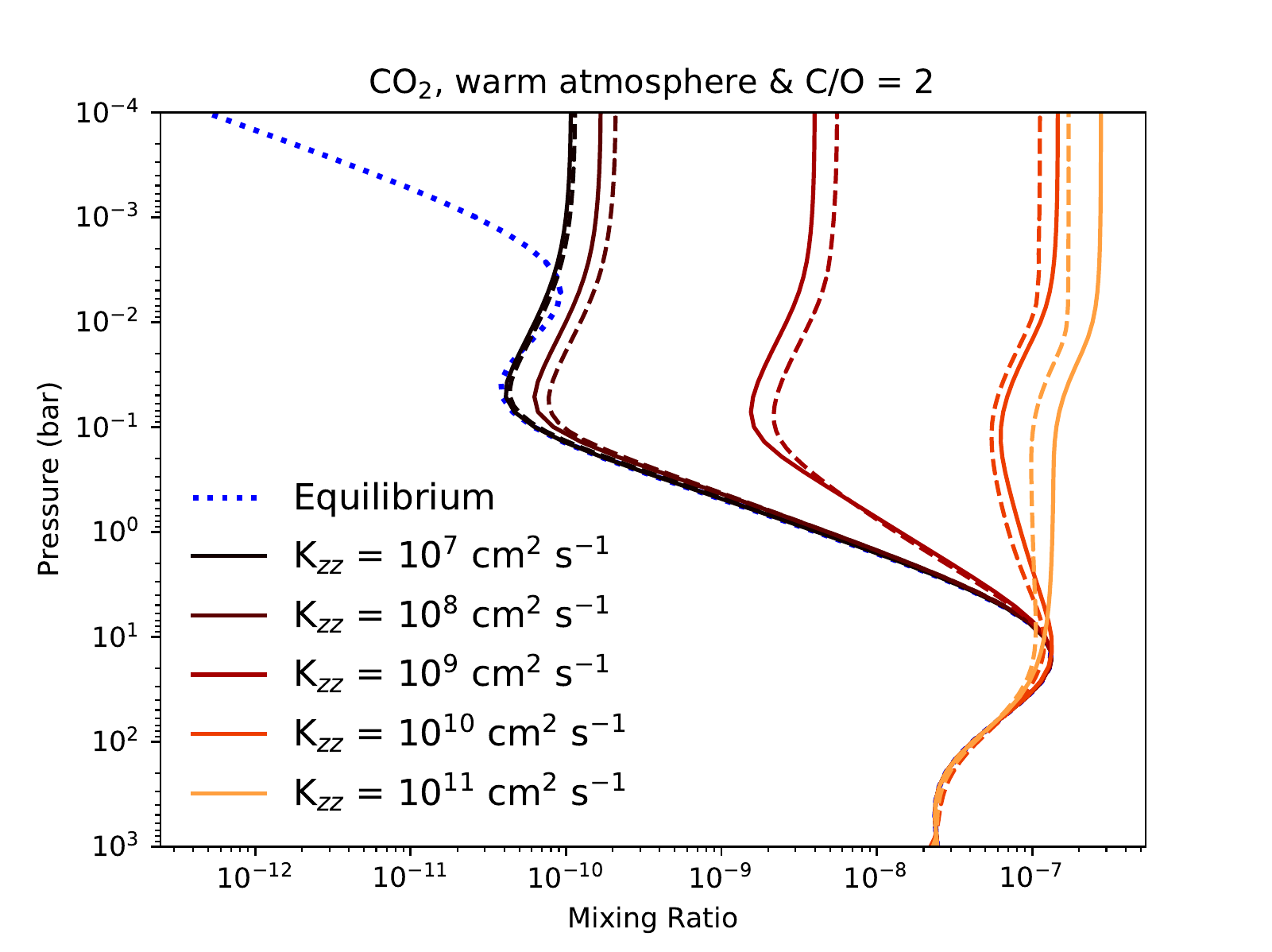}
\end{center}
\caption{Same as Figure \ref{fig:hot_valid} but for the warm atmosphere. For CO, the chemical-relaxation calculations adopting the timescale from \cite{cs06} are shown as dashed-dotted curves. For \ce{CH4}, the mass balance approach using the chemical-relaxation calculations of CO are shown in dashed-dotted curves. For CO$_2$, the chemical-relaxation calculations without considering the coupling to CO and \ce{H2O} are shown in dashed-dotted curves for comparison.}
\label{fig:warm_valid}
\end{figure*}

Figure \ref{fig:warm_valid} shows the mixing ratios of CO, CO$_2$, CH$_4$, H$_2$O and NH$_3$ for a HD 189733b-like atmosphere.  The chemical relaxation method performs with an accuracy of better than a factor of 2 in most parts across a broad range of pressures and mixing ratio values, with \ce{NH3} being least accurate due to the error in estimating its timescale.  For comparison, we show the chemical-relaxation calculations performed using our implementation of the method of \cite{cs06} as dot-dashed curves.  Recall that \cite{cs06} uses essentially a shorter single chemical timescale of CO (tied to a single RLS), whereas the main goal of the present study is to use a set of rate-limiting chemical reactions depending on the temperature and pressure conditions to obtain more accurate timescales. 

\cite{cs06} and \cite{ben18} only calculate CO using the relaxation method and relate \ce{CH4} (and other species) through mass balance, assuming that all the carbons are locked in either \ce{CH4} and CO, hence the mixing ratios $X_{\ce{CH4}}$ and $X_{\ce{CO}}$ are conserved:
\begin{equation}\label{eq:mass_balance}
X_{\ce{CH4}} + X_{\ce{CO}} = \ce{C}/(\ce{H2}+\ce{He})
\end{equation}
where \ce{C}/(\ce{H2}+\ce{He}) is the ratio of carbon atoms to molecular hydrogen and helium (i.e. the bulk gas), determined by solar metallicity. We emphasize that this mass balance relation is only valid when (1) the system is in or close to chemical equilibrium (2) temperature is not too high such that all hydrogen remains in molecular form (3) \ce{CH4} and CO have close abundances. The elemental abundance ($\ce{C}/(\ce{H2}+\ce{He})$ in this case) is a local property, and can be violated by disequilibrium processes\footnotemark[3]. We demonstrate this in the top right panel in Figure \ref{fig:warm_valid}, with the mixing ratios of \ce{CH4} calculated from equation (\ref{eq:mass_balance}) using $X_{\ce{CO}}$ obtained from chemical relaxation (the top left panel in Figure \ref{fig:warm_valid}) in dot-dashed curves. The mass balance approach overpredicts the quenching of \ce{CH4}. Furthermore, as CO is $\sim$3 orders of magnitude more abundant than \ce{CH4} in this case, which requires the estimation of CO to be as accurate as $\sim$3 decimal places for the mass balance approach to work. Unfortunately, this precision is not attainable with the relaxation method or any other kinetics models. Therefore, the mass balance approach can only be applied to a system in chemical equilibrium but not applicable to the relaxation method.

\footnotetext[3]{{For example, CO quenching alone increases the metallicity relative to that in chemical equilibrium, although the effect of changing the metallicity is usually small since 
the quenched species are trace gases.}}


For CO$_2$, we show for comparison the calculation of relaxing CO$_2$ to the equilibrium abundance of CO2 (dot-dashed curves) versus to its pseudo-equilibrium value as determined by the quenched abundances of CO and H$_2$O (solid curves).  The difference in accuracy is substantial: $\sim 10\%$ versus an order of magnitude.  Neglecting this effect will lead to the inaccurate prediction that the mixing ratio of CO$_2$ is close to equilibrium.

For illustration, we show the mixing ratios of CH$_4$ when the metallicity is $100 \times$ solar, as well as those of H$_2$O and CO$_2$ when $\mbox{C/O}=2$.  In the latter case, H$_2$O loses its dominance to CH$_4$ and its mixing ratio becomes sensitive to the strength of vertical mixing.  In these cases, our more general treatment of chemical relaxation allows the accuracy of the method to remain about the same as for the solar-metallicity case.

\subsection{Cool atmospheres (GJ 1214b-like)}
 
\begin{figure*}
\begin{center}
\vspace{-0.1in}
\includegraphics[width= 0.87\columnwidth]{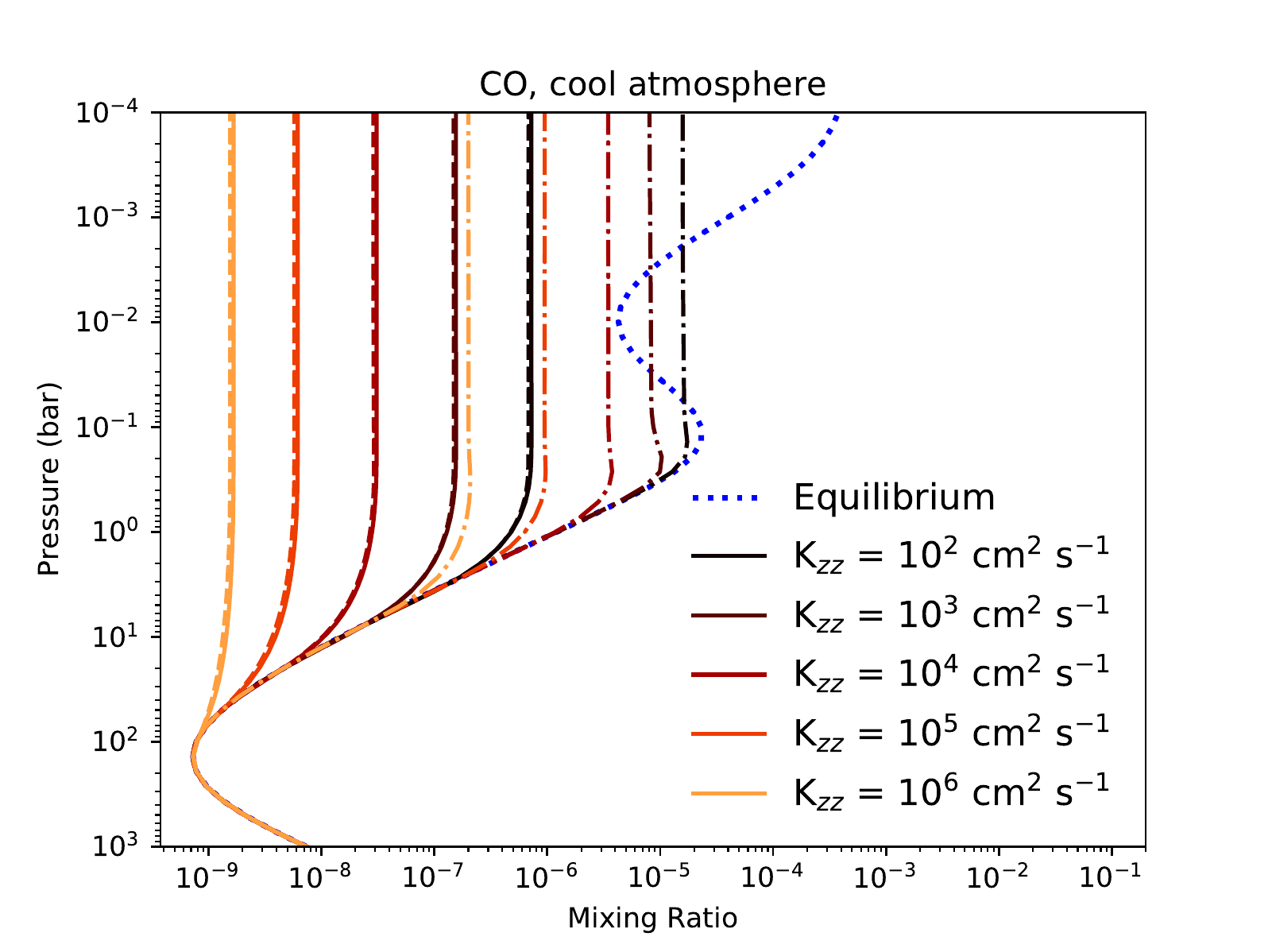}
\includegraphics[width= 0.87\columnwidth]{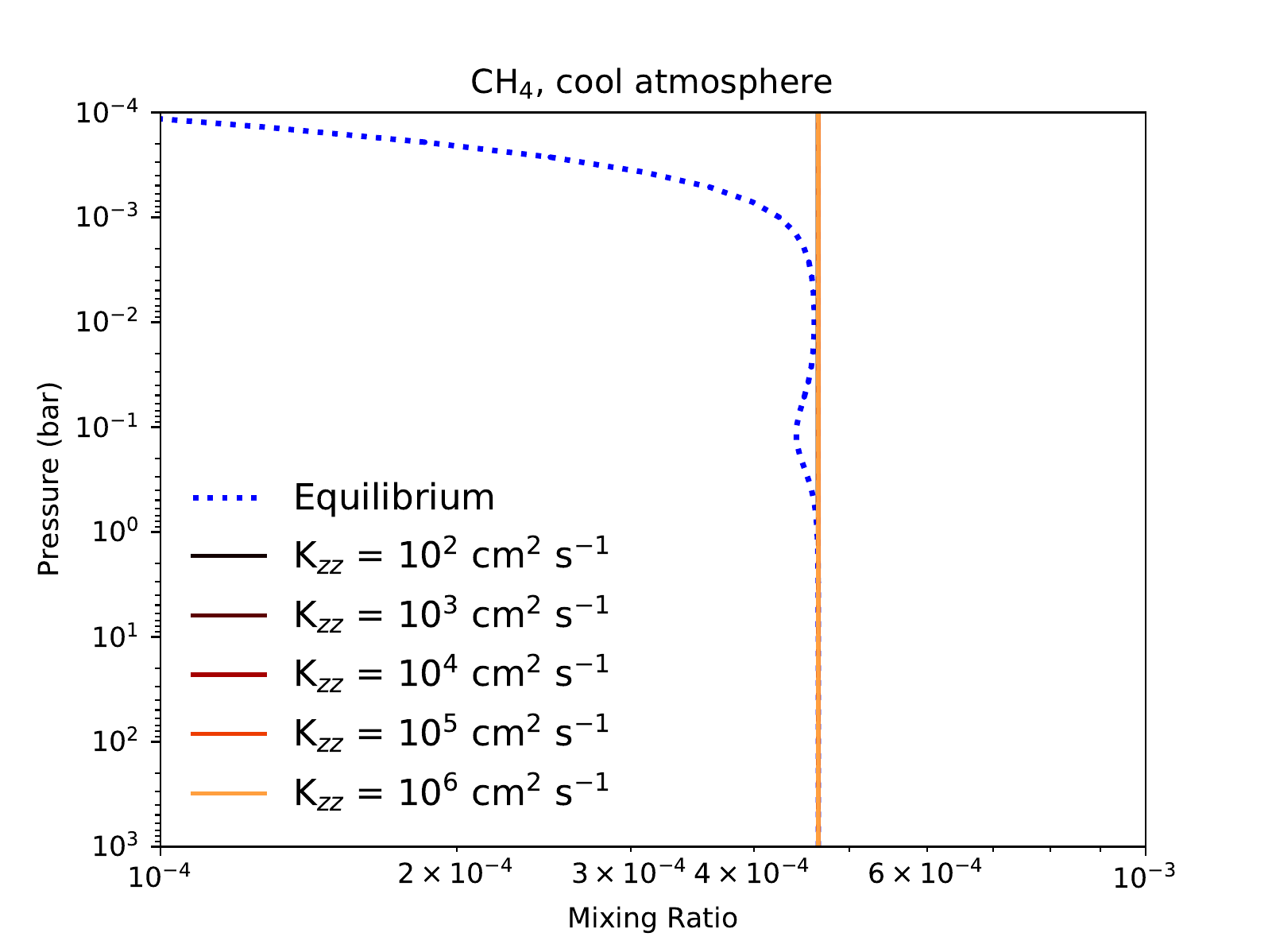}
\includegraphics[width= 0.87\columnwidth]{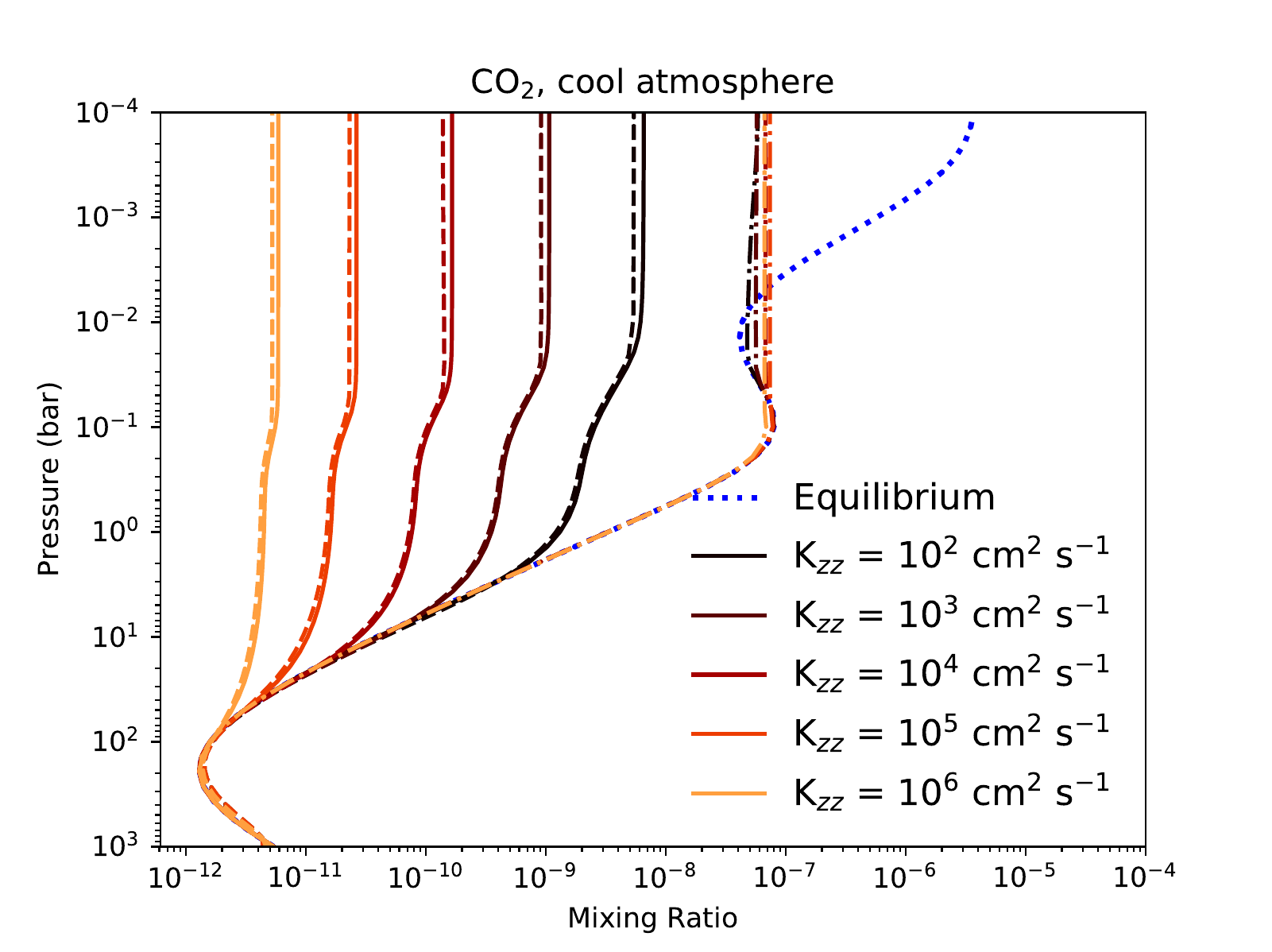}
\includegraphics[width= 0.87\columnwidth]{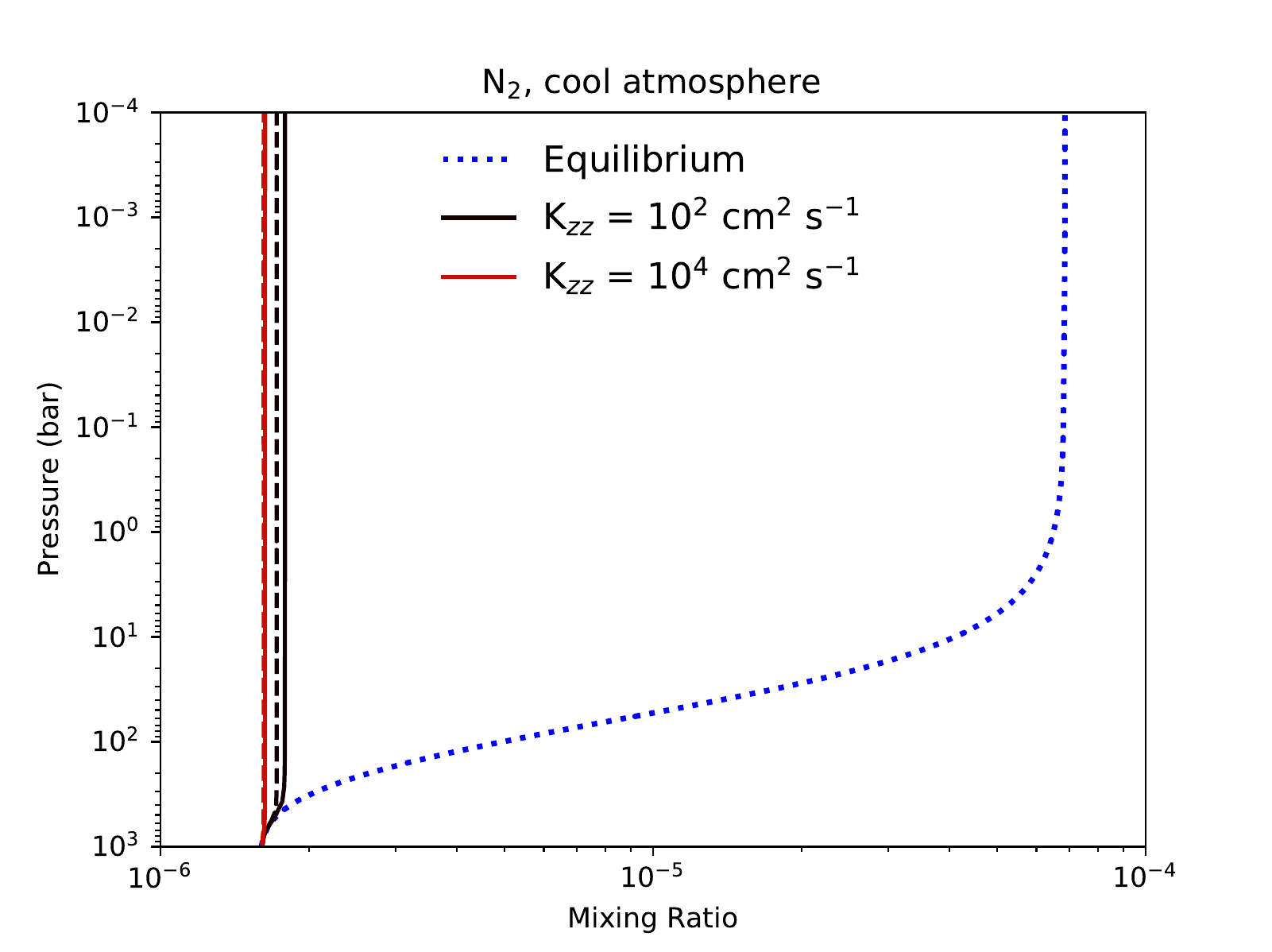}
\end{center}
\caption{Same as Figure \ref{fig:hot_valid} but for the cool atmosphere. For CO, the chemical-relaxation calculations adopting the timescale from \cite{cs06} are shown as dashed-dotted curves. For CO$_2$, the chemical-relaxation calculations without considering the coupling to CO and \ce{H2O} are shown in dashed-dotted curves for comparison.}
\label{fig:cool_valid}
\end{figure*}

Figure \ref{fig:cool_valid} shows the mixing ratios of CO, CO$_2$, CH$_4$ and N$_2$ for a GJ 1214b-like atmosphere.  In this range of temperatures, the chemical relaxation method is highly accurate ($\sim 10\%$). For CO, the difference from using the single-RLS timescale of \cite{cs06} increases significantly to a few orders of magnitudes. However, we note that photochemistry will potentially influence the abundances in these cooler atmospheres, which is not taken into account in this work.


\section{Summary and discussion}
\label{sec:summary}

\subsection{Summary}

Inspired by the pioneering work of \cite{cs06}, we have revisited the chemical relaxation method, which seeks to greatly enhance computational efficiency by replacing the network in a chemical kinetics calculation with a few independent source/sink terms that ``relax" towards chemical equilibrium on a prescribed timescale.  There is a precedent of using relaxation methods as a substitute for radiative transfer, where the timescale is then associated with radiative cooling (e.g., \citealt{hs94}).  The main lessons learned from our study are:
\begin{itemize}

\item The rate-limiting reaction that determines the chemical timescale depends on the temperature and pressure.  For \ce{CH4}-CO and \ce{NH3}-\ce{N2}-HCN interconversion, we show across a broad range in temperature and pressure (500--3000 K, 0.1 mbar to 1 kbar) that there are multiple rate-limiting reactions, and that the chemical timescale cannot be easily fitted by an Arrhenius-like function.

\item By comparing full chemical-kinetics to chemical-relaxation calculations in one dimension, we show that the latter is accurate to within an order of magnitude for WASP-18b-like atmospheres, $\sim$ a factor of 2 for HD 189733b-like atmospheres and $\sim 10\%$ for GJ 1214-b-like atmospheres.  Essentially, the chemical relaxation method is more accurate when the species either fully quench or retain chemical equilibrium. The discrepancies become larger when the species do not fully quench, because the behavior is more sensitive to the timescale in this situation. Overall, species at lower temperatures tend to fully quench (e.g., \citealt{moses16}) since the timescale of the main quenched species (CO and \ce{N2}) quickly increases with altitude (see Figure \ref{fig:tau_contours}). This bodes well as the currently characterizable atmospheres will become cooler as observational methods advance, but the effects of photochemistry need to be examined.

\item The relaxation method increases the computational speed by at least 100 times compared to running a full kinetics model. More importantly, the relaxation method allows decoupling from the chemical network. Only the species of interest need to be included, which will significantly ease the burden of adding numerous tracers to the dynamical core.


\end{itemize}

\subsection{Opportunities for future work}

There are ample opportunities for future work. We have recently finished and submitted the work of coupling our chemical relaxation method to our GCM \citep{mendonca16} and studying the interaction between atmospheric dynamics and chemistry in three dimensions.  It remains an open question if photochemistry may be reasonably approximated by chemical relaxation, since it will require a pre-calculated photochemical steady state for a given atmospheric condition and stellar flux. Generalizing chemical relaxation to work in the regime of atmospheres with Earth-like temperatures will be useful as we march towards the study of exo-climates similar to our own.

\acknowledgments
S.-M.T. and K.H. acknowledge partial financial support from the PlanetS National Center of Competence in Research (NCCR), the Center for Space and Habitability, the Swiss National Science Foundation and the Swiss-based MERAC Foundation, as well as useful conversations with Julie Moses and Paul Rimmer.

\software{\\
Python\footnotemark[4], SciPy\footnotemark[5], NumPy\footnotemark[6] \citep{numpy}, Matplotlib\footnotemark[7] \citep{matplotlib}}
\footnotetext[4]{\url{http://www.python.org}}
\footnotetext[5]{\url{http://scipy.org}}
\footnotetext[6]{\url{http://numpy.org}}
\footnotetext[7]{\url{http://matplotlib.org}}

\appendix
\section{Full expressions for the chemical timescales}
\label{app:f_tau}
[X] represents the number density of species X in chemical equilibrium and M refers to any third body.\\

\footnotesize
For C/O $<=$ 1:

\begin{equation}
\begin{cases}
\begin{aligned}
& \tau_{\ce{CH4}} = \frac{[\ce{CH4}]}{k_1[\ce{CH3}][\ce{OH}][\ce{M}] + \mathrm{max}(\mathrm{min}(k_2[\ce{CH2OH}][\ce{H}], k_3[\ce{CH2OH}][\ce{M}]), k_4[\ce{CH3}][\ce{O}] + k_9[\ce{CH3OH}][\ce{H}] + \mathrm{min}(k_5[\ce{C}][\ce{OH}], k_6[\ce{CH4}][\ce{H}])} + \tau_{\rm H_2}\times \frac{3 [{\rm CO}]}{[{\rm H_2}]}\\
& \tau_{\rm H_2} =  \frac{[\ce{H2}]}{k_{\ce{H}}[\ce{H}][\ce{H}][\ce{M}]}\\
\end{aligned}
\end{cases}
\end{equation}

\begin{equation}
\tau_{\ce{CO}} = \frac{[\ce{CO}]}{k_1[\ce{CH3}][\ce{OH}][\ce{M}] + \mathrm{max}(\mathrm{min}(k_2[\ce{CH2OH}][\ce{H}], k_3[\ce{CH2OH}][\ce{M}]), k_4[\ce{CH3}][\ce{O}]   + k_9[\ce{CH3OH}][\ce{H}] + \mathrm{min}(k_5[\ce{C}][\ce{OH}], k_6[\ce{CH4}][\ce{H}])} + \tau_{\rm H_2}\times \frac{3 [{\rm CO}]}{[{\rm H_2}]}\\
\end{equation}

\begin{equation}
\tau_{\ce{H2O}} = \frac{[\ce{H2O}]}{k_1[\ce{CH3}][\ce{OH}][\ce{M}] + \mathrm{max}(\mathrm{min}(k_2[\ce{CH2OH}][\ce{H}], k_3[\ce{CH2OH}][\ce{M}]), k_4[\ce{CH3}][\ce{O}] + k_9[\ce{CH3OH}][\ce{H}] + \mathrm{min}(k_5[\ce{C}][\ce{OH}], k_6[\ce{CH4}][\ce{H}])} + \tau_{\rm H_2}\times \frac{3 [{\rm CO}]}{[{\rm H_2}]}\\
\end{equation}

For C/O $>$ 1:

\begin{equation}
\tau_{\ce{CH4}} = \frac{[\ce{CH4}]}{k_1[\ce{CH3}][\ce{OH}][\ce{M}] + \mathrm{min}(k_2[\ce{CH2OH}][\ce{H}], k_3[\ce{CH2OH}][\ce{M}]) + k_9[\ce{CH3OH}][\ce{H}] + 
\mathrm{max}(k_8[\ce{C2H2}][\ce{O}], k_{10}[\ce{C2H2}][\ce{OH}])}
\end{equation}

\begin{equation}
\tau_{\ce{CO}} = \frac{[\ce{CO}]}{k_1[\ce{CH3}][\ce{OH}][\ce{M}] + \mathrm{min}(k_2[\ce{CH2OH}][\ce{H}], k_3[\ce{CH2OH}][\ce{M}]) + k_9[\ce{CH3OH}][\ce{H}] + 
\mathrm{max}(k_8[\ce{C2H2}][\ce{O}], k_{10}[\ce{C2H2}][\ce{OH}])}
\end{equation}

\begin{equation}
\tau_{\ce{H2O}} = \frac{[\ce{H2O}]}{k_1[\ce{CH3}][\ce{OH}][\ce{M}] + \mathrm{min}(k_2[\ce{CH2OH}][\ce{H}], k_3[\ce{CH2OH}][\ce{M}]) + k_9[\ce{CH3OH}][\ce{H}] + 
\mathrm{max}(k_8[\ce{C2H2}][\ce{O}], k_{10}[\ce{C2H2}][\ce{OH}])}
\end{equation}
\\
\hrule

\begin{equation}
\begin{aligned}\label{eq:co2_pseudo}
& \\
& [\ce{CO2}]_{\mbox{pseudo-eq}} = \frac{[\ce{CO}][\ce{H2O}][\ce{H}]_{\mbox{eq}}}{[\ce{CO}]_{\mbox{eq}}[\ce{H2O}]_{\mbox{eq}}[\ce{H}]}\\
& \tau_{\ce{CO2}} = \frac{[\mbox{CO}_2]}{k_{\mbox{CO}_2}[\mbox{CO}][\mbox{OH}]}
\end{aligned}
\end{equation}

\begin{equation}
\\
\tau_{\ce{NH3}} = \frac{1}{2} \left( \frac{[\ce{NH3}]}{ \mathrm{max}(k_{11}[\ce{NH2}][\ce{NH2}] , k_{12}[\ce{N2H3}][\ce{M}]) + k_{13}[\ce{NH}][\ce{NH2}] + k_{14}[\ce{NO}][\ce{NH2}]) + k_{15}[\ce{N}][\ce{NO}]} + \tau_{\rm H_2}\times \frac{3 [{\rm N2}]}{[{\rm H_2}] } \right) 
\end{equation}

\begin{equation}
\tau_{\ce{N2}} = \frac{[\ce{N2}]}{ \mathrm{max}(k_{11}[\ce{NH2}][\ce{NH2}] , k_{12}[\ce{N2H3}][\ce{M}]) + k_{13}[\ce{NH}][\ce{NH2}] + k_{14}[\ce{NO}][\ce{NH2}]) + k_{15}[\ce{N}][\ce{NO}]} + \tau_{\rm H_2}\times \frac{3 [{\rm N2}]}{[{\rm H_2}] }
\end{equation}

\normalsize

\section{Chemical pathway analysis}
\label{app:pathway}
Finding the chemical pathway is similar to the path finding problem in graph theory. That is,
all species in the network are presented by nodes and reactions between them form the edges, weighted by the reaction rates (faster reactions form shorter connections). It is equivalent to finding the least time-consuming route from a a starting node to an end node in the network and the total time-cost of the route can be approximated by the slowest edge (the rate-limiting step). 
While there are several different algorithms, e.g., \cite{ralph} identifies the dominant pathways
that are most efficient in removing/producing a species of interest from kinetics results with temporal evolution. We are particularly interested in finding the pathway in chemical equilibrium, which fulfills the need for estimating the chemical timescales using equilibrium abundances. For this purpose, We implement Dijkstra's algorithm \citep{Dji}, which is easy to implement and highly efficient in finding fastest paths in the network problems \citep{Viswanath}. 

Examples of comparing different chemical networks by identifying the pathways with associated rate-limiting steps are shown in Figures \ref{fig:path_tool} and \ref{fig:path_tool2}. 

Below are the steps used in Dijkstra's algorithm to find the shortest path.\\
1. Create a list of visited nodes (initially empty). Assign tentative distance values to all nodes: set it to zero for the initial node and to infinity for all other nodes. Set the initial node as the current node.\\
2. From the current node, consider all of its neighbors and calculate their tentative distances.
Update the distance if the new value is smaller than the previously assigned value.\\
3. Include the current node in the visited-node list. A visited node will never be checked again.\\
4. Stop if the destination node has been marked visited. The pathway has been found and the longest edge (the slowest step) in the path is the rate-limiting step. Otherwise, select the unvisited node with the smallest tentative distance, set it as the new ``current node", and go back to step 2.\\

\begin{figure}
\begin{minipage}{0.4\columnwidth}
\begin{center}
\includegraphics[width= \columnwidth]{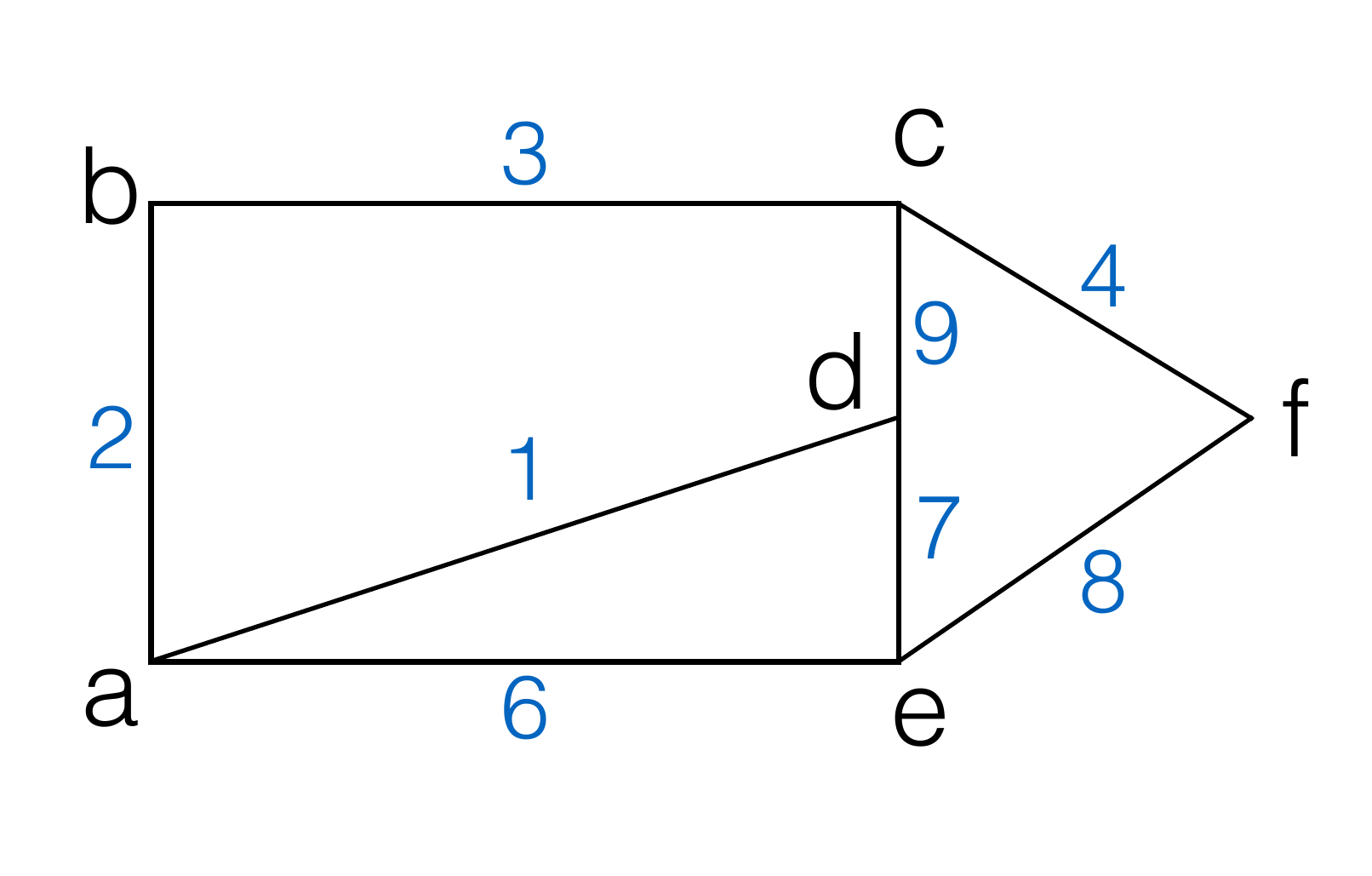}
\end{center}
\vspace{+0.3in}
\caption{Finding the shortest path from a to f using Dijkstra's algorithm. There are six nodes labeled a to f connected by the edges labeled with distance. With the application to chemical networks, the nodes represent species and the edges are reactions with different rates.}
\label{fig:Dij_ex}
\end{minipage}
\hfill
\begin{minipage}{0.4\columnwidth}
\begin{center}
\vspace{-0.15in}
\includegraphics[width= \columnwidth]{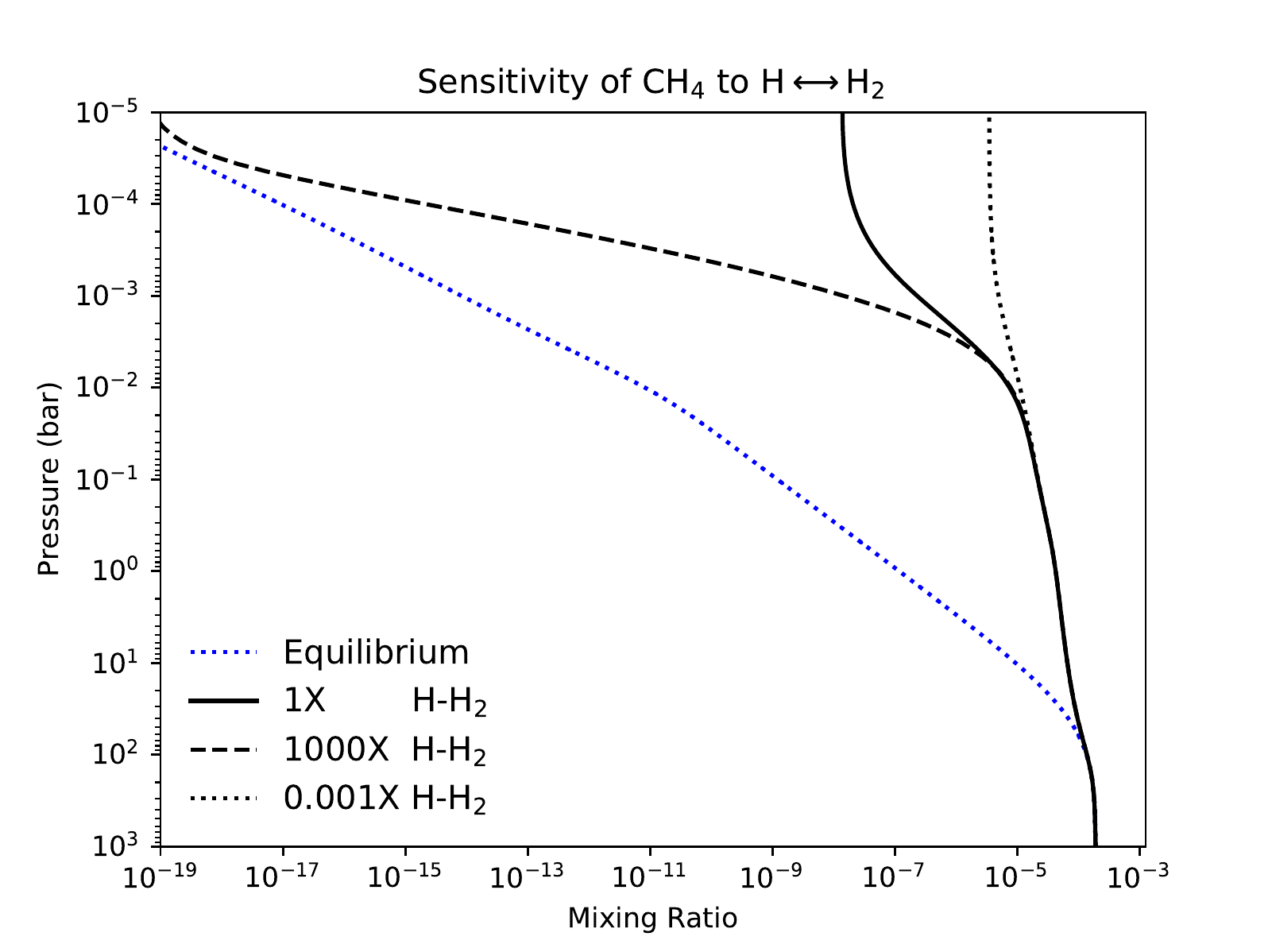}
\end{center}
\caption{A sensitivity test showing how the quenching of \ce{CH4} changes when we artificially increases/decrease the rate of H-H$_2$ dissociation/recombination by 1000 times.}
\label{fig:sensitivity}
\end{minipage}
\end{figure}

\begin{figure}
\begin{center}
\vspace{-0.1in}
\includegraphics[width= 0.25\columnwidth]{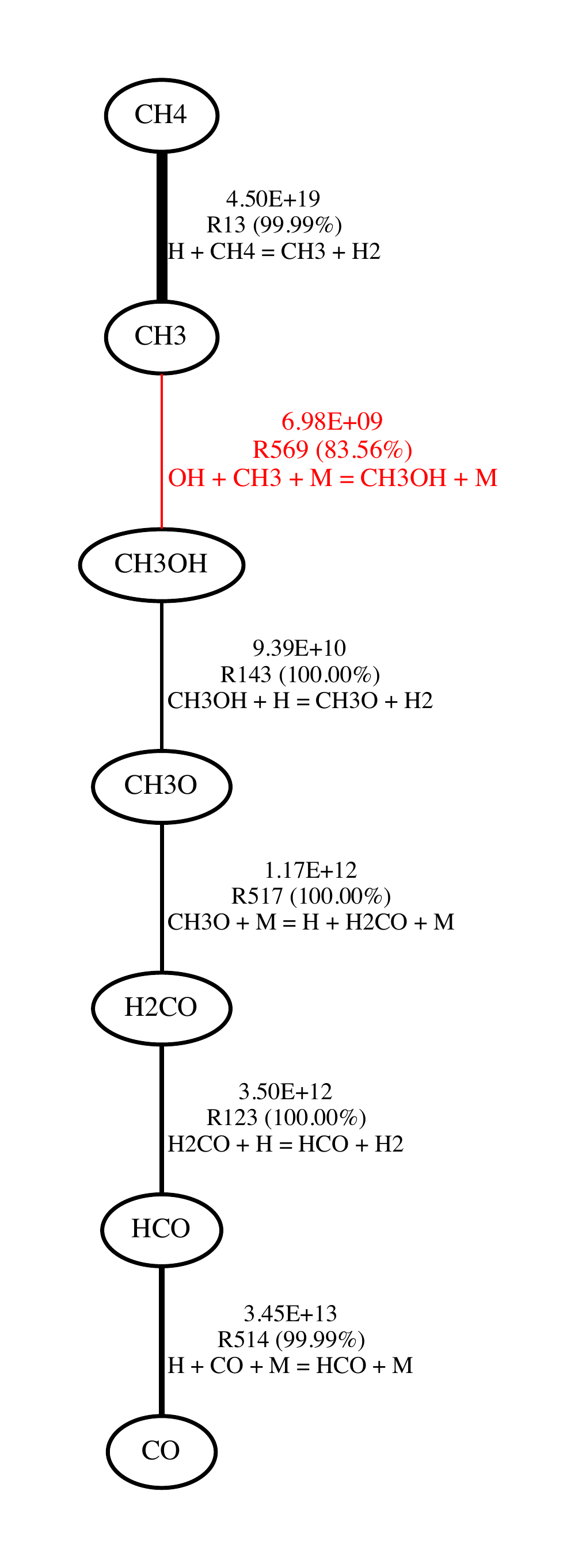}
\includegraphics[width= 0.25\columnwidth]{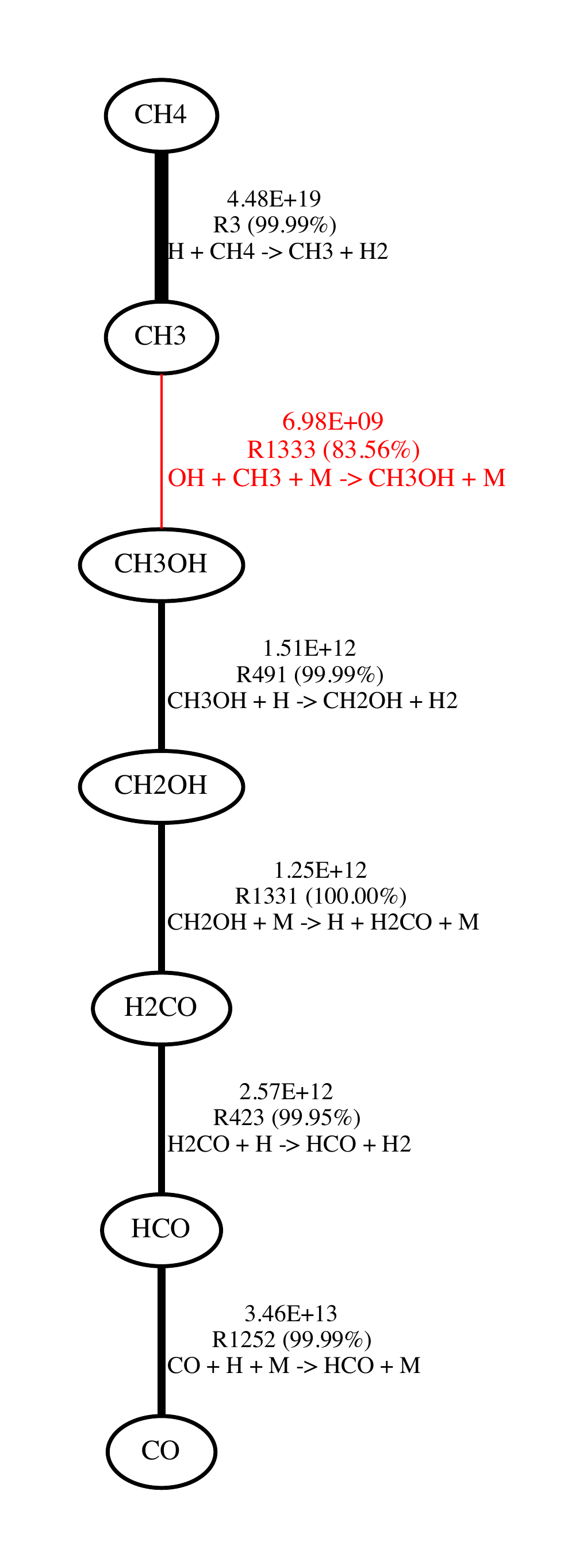}
\includegraphics[width= 0.25\columnwidth]{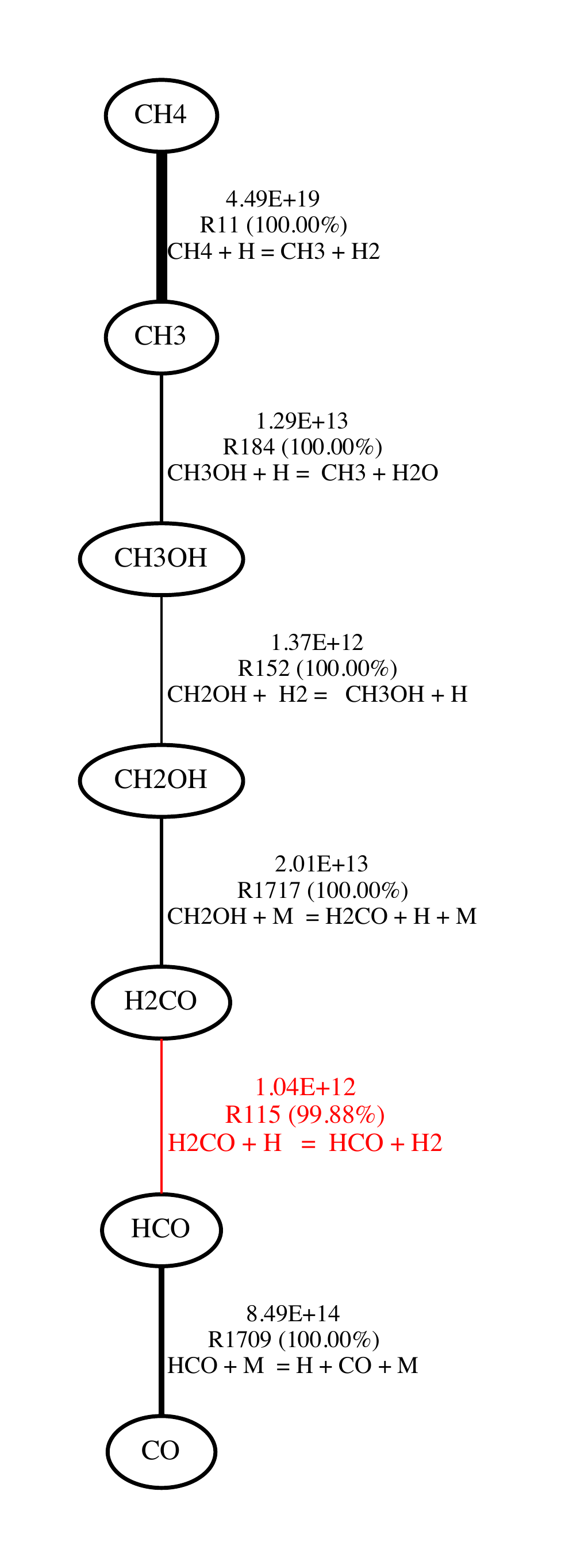}
\end{center}
\vspace{-0.2in}
\caption{More examples of \ce{CH4}-\ce{CO} pathway analysis at $T = 1200$ K and $P = 500$ bar with the chemical network from VULCAN (left), \cite{moses11} (middle), and \cite{venot12} (right). Thicker lines represent faster reaction rates (denoted by the first-row numbers shown in in cm$^{-3}$s$^{-1}$ and the percentage of contribution to the interconversion rate is also provided is also provided) and the red lines are the rate-limiting steps. The selected temperature and pressure values are based on the CO quenched level in the deep atmosphere of Jupiter to identify the different chemical pathways as pointed out in Figure 17 of \cite{wang16}.}
\label{fig:path_tool2}
\end{figure}

We demonstrate the steps in the following example:\\
Consider the simple network with nodes a-f in Figure \ref{fig:Dij_ex}. We will try to find the shortest path between a and f.\\ \\
- The tentative distance values are assigned to a,b,c,d,e,f as [0, inf, inf, inf, inf, inf]. The current node is a.\\
- The adjacent nodes are b, d, and e. The distance values are updated to [0, 2, inf, 1, 6, inf].\\
- The list of visited nodes is now updated to [a]. d as the unvisited node with the smallest distance value becomes the current node.\\
- The adjacent nodes are c and e. The distance values are updated to [0, 2, 10, 1, 6, inf].\\
- The list of visited nodes is now updated to [a, d]. b as the unvisited node with the smallest distance value becomes the current node.\\
- We repeat the above steps (second to fifth steps in the algorithm) until f has been included in the list of visited nodes. We then have the distance values as [0, 2, 5, 1, 6, 9]. The shortest path is obtained by going back from f and following the smallest distance value. The path is a $\rightarrow$ b $\rightarrow$ c $\rightarrow$ f.


\begin{table}[!htb]
\begin{center}
\caption{Rate-limiting reactions/steps and their rate coefficients. $k_0$ and $k_{\infty}$ are the low-pressure and high-pressure limiting rate coefficients, respectively. The rate coefficient for the termolecular or thermal dissociation reaction is expressed by equation (28) in \cite{tsai17}. Rev indicates the reverse reactions of Rx, which are used to derive the rate constants of Rx using the equilibrium constant calculated by the NASA polynomials $^8$, as described in Appendix E in \cite{tsai17}.}
\label{tab: rate_const}
\begin{tabular}{llll}
\hline
Index & Reaction & Rate Coefficient & Reference \\
\hline
R1 & \ce{ OH + CH3 ->T[M] CH3OH} & \hspace*{-0.8cm}\makecell[l]{$k_0$  = 1.93 $\times 10^{3}$ $T^{-9.88}$ $\exp(-7544/T) + $\\5.11$\times 10^{-11}$ $T^{-6.25}$ $\exp(-1433/T)$\\$k_{\infty}$ = 1.03 $\times 10^{-10}$ $T^{-0.018}$ $\exp(16.74/T)$} & \cite{moses11}\\
R2$_{\mbox{rev}}$ & \ce{ CH2OH + H -> OH + CH3 } & 1.60 $ \times 10^{-10}$ & NIST $^9$ 1987TSA471\\
R3 & \ce{ CH2OH ->T[M] H + H2CO }  & \hspace*{-0.8cm}\makecell[l]{$k_0$  = 1.66 $ \times 10^{-10}$ $\exp(-12630/T)$\\$k_{\infty}$ = 3 $ \times 10^{9}$ $\exp(-14600/T)$} & \hspace*{-0.8cm}\makecell[l]{NIST 1987TSA471\\NIST 1975BOW343}\\
R4 & \ce{ CH3 + O -> H2CO + H } & 1.4 $ \times 10^{-10}$ & NIST 1992BAU/COB411-429\\
R5 & \ce{ OH + C -> CO + H } & 1.05 $\times 10^{-12}$ $ T^{0.5} $ & NSRDS 67\\
R6 & \ce{ H + CH4 -> CH3 + H2 } & 2.20 $ \times 10^{-20}$ $ T^{3} $  $\exp(-4040/T)$ & NIST 1992BAU/COB411-429\\
R7 & \ce{ CH3OH + H -> CH3O + H2 } & 6.82 $ \times 10^{-20}$ $ T^{2.685} $ $\exp(-4643/T)$ &NIST 1984WAR197C\\
R8 & \ce{ C2H2 + O -> CH2 + CO  } & 6.78 $ \times 10^{-16}$ $ T^{1.5} $  $\exp(-854/T)$ & NIST 1987CVE261\\
R9$_{\mbox{rev}}$ & \ce{ CH3OH + H -> CH3 + H2O } & 4.91 $ \times 10^{-19}$ $ T^{2.485} $  $\exp(-10380/T)$ & \cite{moses11}\\
R10 & \ce{ C2H2 + OH -> CH3 + CO } & 8.04 $ \times 10^{-28}$ $ T^{4} $  $\exp(1010/T)$ & NIST 1989MIL/MEL1031-1039\\
R11 & \ce{ NH2 + NH2 -> N2H2 + H2 } & 2.89 $ \times 10^{-16}$ $ T^{1.02} $  $\exp(-5930/T)$ & NIST 2009KLI/HAR10241-10259\\
R12 & \ce{N2H3 + ->T[M] N2H2 + H} & \hspace*{-0.8cm}\makecell[l]{$k_0$  = 3.49 $ \times 10^{38}$ $T^{-13.13}$ $\exp(-36825/T)$\\$k_{\infty}$ = 7.95 $ \times 10^{13}$ $\exp(-27463/T)$} & \cite{hm03}\\
R13 & \ce{ NH + NH2 -> N2H2 + H } & 6.98 $ \times 10^{-10}$ $ T^{-0.27} $  $\exp(39/T)$ & NIST 2009KLI/HAR10241-10259\\ 
R14 & \ce{ NO + NH2 -> N2 + H2O } & 7.9 $ \times 10^{-9}$ $ T^{-1.1} $  $\exp(-98/T)$ & NIST 1994DIA/YU4034-4042\\
R15 & \ce{  N + NO -> N2 + O } & 3.7 $ \times 10^{-11}$ & NIST 1992MIC/LIM3228-3234\\
R16 & \ce{N2H4 + H -> N2H3 + H2} & 1.17 $ \times 10^{-11}$	$\exp(-1260/T)$ & NIST 1995VAG777-790\\
$k_{\ce{H}}$ & \ce{ H + H ->T[M] H2}  & \hspace*{-0.8cm}\makecell[l]{$k_0$  = 2.7 $ \times 10^{-31}$ $ T^{-0.6} $\\$k_{\infty}$ = 3.31 $ \times 10^{-6}$ $ T^{-1} $} & \hspace*{-0.8cm}\makecell[l]{NIST 1992BAU/COB411-429\\NIST 1965JAC/GIE3688}\\
$k_{\ce{CO2}}$ & \ce{CO + OH -> CO2 +H } & 1.05 $ \times 10^{-17}$ $ T^{1.5} $  $\exp(259/T)$ & NIST 1992BAU/COB411-429\\
\end{tabular}
\end{center}
\end{table}
\footnotetext[8]{\url{http://garfield.chem.elte.hu/Burcat/burcat.html}}
\footnotetext[9]{Shown by Squib in the NIST database}


\label{lastpage}

\end{document}